\documentclass[twocolumn,english,aps,pra,superscriptaddress,floatfix]{revtex4-1}
\usepackage[T1]{fontenc}
\usepackage[latin9]{inputenc}
\usepackage{color}
\usepackage{bm}
\usepackage{bbm}
\usepackage{dsfont}
\usepackage{xcolor}
\usepackage{amsmath}
\usepackage{amssymb}
\usepackage{graphicx}
\usepackage{natbib}
\usepackage{braket}
\usepackage{psfrag}
\usepackage{tikz}
\usepackage[normalem]{ulem}
\usepackage{qcircuit}
\usepackage{soul}
\usepackage{mathtools}
\usepackage{afterpage}
\usepackage{algorithm}
\usepackage{float}
\usepackage{algpseudocode}
\usepackage{comment}

\usepackage[colorlinks=true,urlcolor=blue,citecolor=blue,linkcolor=blue]{hyperref}

\newcommand{\be}{\begin{equation}}
\newcommand{\ee}{\end{equation}}
\newcommand{\bq}{\begin{eqnarray}}
\newcommand{\eq}{\end{eqnarray}}
\newcommand{\bea}{\begin{eqnarray}}
\newcommand{\eea}{\end{eqnarray}}
\newcommand{\ba}{\begin{align}}
\newcommand{\ea}{\end{align}}

\newcommand{\proj}[1]{\ket{#1}\bra{#1}}

\newcommand{\avr}[1]{\left \langle#1 \right \rangle}

\newcommand{\tr}[1]{{\rm tr}\left[{#1}\right]}

\newcommand{\bR}{\mathbbm{R}}

\newcommand{\bC}{\mathbbm{C}}
\newcommand{\bZ}{\mathbbm{Z}}

\newcommand{\half}{\frac{1}{2}}

\newcommand{\myVec}[1]{{\bm{#1}}}

\def\qed{\leavevmode\unskip\penalty9999 \hbox{}\nobreak\hfill
    \quad\hbox{\leavevmode  \hbox to.77778em{%
               \hfil\vrule   \vbox to.675em%
               {\hrule width.6em\vfil\hrule}\vrule\hfil}}
     \par\vskip3pt}

\begin{document}

\title{Covariant quantum kernels for data with group structure}

\author{Jennifer R. Glick}
\affiliation{IBM Quantum, IBM T.J. Watson Research Center, Yorktown Heights, NY 10598, USA}
\author{Tanvi P. Gujarati}
\affiliation{IBM Quantum, Almaden Research Center, San Jose, California 95120, USA}
\author{Antonio D. C\'orcoles}
\affiliation{IBM Quantum, IBM T.J. Watson Research Center, Yorktown Heights, NY 10598, USA}
\author{\mbox{Youngseok  Kim}}
\affiliation{IBM Quantum, IBM T.J. Watson Research Center, Yorktown Heights, NY 10598, USA}
\author{Abhinav Kandala}
\affiliation{IBM Quantum, IBM T.J. Watson Research Center, Yorktown Heights, NY 10598, USA}
\author{Jay M. Gambetta}
\affiliation{IBM Quantum, IBM T.J. Watson Research Center, Yorktown Heights, NY 10598, USA}
\author{Kristan Temme}
\email{kptemme@ibm.com}
\affiliation{IBM Quantum, IBM T.J. Watson Research Center, Yorktown Heights, NY 10598, USA}

\date{\today}

\begin{abstract}
The use of kernel functions is a common technique to extract important features from data sets. A quantum computer can be used to estimate kernel entries as transition amplitudes of unitary circuits. Quantum kernels exist that, subject to computational hardness assumptions, cannot be computed classically. It is an important challenge to find quantum kernels that provide an advantage in the classification of real-world data. We introduce a class of quantum kernels that can be used for data with a group structure. The kernel is defined in terms of a unitary representation of the group and a fiducial state that can be optimized using a technique called kernel alignment. We apply this method to a learning problem on a coset-space that embodies the structure of many essential learning problems on groups. We implement the learning algorithm with $27$ qubits on a superconducting processor.
\end{abstract}

\maketitle

The core tenet of the kernel method in machine learning is that it allows one to apply linear statistical methods to data sets that are complex and non-linear in nature. The kernel function $K$ corresponds to an inner product of vectors in a (potentially) high-dimensional Euclidean space referred to as the feature space \cite{boser1992training}. The datum $\myVec{x} \in {\cal X}$ is mapped to this high-dimensional space by means of a non-linear feature map $\Phi(\myVec{x})$. This feature map has to be chosen in such a way that the data, initially not tractable by linear methods, can be linearly separated in the higher-dimensional space. The choice of the non-linear map is therefore central to this approach. The use of the kernel trick allows one to process the data in the high-dimensional space without explicitly computing the feature vector. This trick has found its way into many machine learning tasks such as classification~\cite{boser1992training}, regression~\cite{vapnik1996regression, smola2004tutorial}, clustering~\cite{benhur2002support, girolami2002mercer}, correlation analysis~\cite{lai2000kernel} and filtering~\cite{liu2011kernel}. The most prominent application of the kernel method is arguably for binary classification in the use of the support vector machine (SVM)~\cite{boser1992training, vapnik2013nature}. This is also the learning problem we consider in the experimental implementation here. After seeing $m$ training samples $\myVec{x}_i \in {\cal X}$ with labels $y_i = \pm 1$, we train a classifier $f$ that accurately predicts the label $y = f(\myVec{x})$ of a previously unseen datum $\myVec{x}$. 

A quantum computer can be used to perform the feature mapping into a quantum-enhanced feature space and estimate the kernel matrix for the training data \cite{havlivcek2019supervised, schuld2019quantum}. This quantum kernel can then be used in most machine learning algorithms that use the kernel method. In fact, it was observed in \cite{havlivcek2019supervised} that many of the recently introduced variational quantum algorithms ~\cite{mitarai2018quantum, farhi2018classification, grant2018hierarchical, schuld2020circuit, benedetti2019parameterized} reduce to a quantum kernel method, since these algorithms are only linear methods in the quantum feature space. Following \cite{havlivcek2019supervised}, a datum $\myVec{x} \in {\cal X}$ is mapped to an $n$-qubit quantum feature state $\Phi(\myVec{x}) = U(\myVec{x})\proj{0^n}U ^\dagger(\myVec{x})$ through a unitary circuit family $U(\myVec{x})$. The unitary depends non-linearly on the datum and needs to have an efficient implementation. The kernel entry for two samples $\myVec{x},\tilde{\myVec{x}}$ is obtained as the Hilbert-Schmidt inner product $K(\myVec{x},\tilde{\myVec{x}}) = \tr{\Phi^\dagger(\myVec{x}) \Phi(\tilde{\myVec{x}})}$ of the two quantum feature states, and can be understood as the transition amplitude
\begin{equation}\label{eqn:kernel-definition}
    K(\myVec{x},\tilde{\myVec{x}}) = |\bra{0^n} U^\dagger(\myVec{x}) U(\tilde{\myVec{x}}) \ket{0^n}|^2.
\end{equation}
The kernel entry can be estimated on a quantum computer by evolving the initial state $\ket{0^n}$ with  $U^\dagger(\myVec{x}) U(\tilde{\myVec{x}})$ and recording the frequency of the all-zero outcome $0^n$. This procedure (c.f. Appendix A.1) is referred to as quantum kernel estimation (QKE). 

Learning algorithms that use QKE have a proven advantage over all classical learners for specifically constructed learning problems \cite{liu2020rigorous}. A core challenge is to establish this advantage in practically relevant settings. We take steps to address this challenge by identifying a class of learning problems that provide a natural fit for QKE and generalize the result in \cite{liu2020rigorous}. What these learning problems have in common is that the data space is a subset of a group ${\cal X} \subseteq G$. The study of data with group structure has a long tradition in statistics \cite{diaconis1988group}. Important learning problems such as ranking \cite{diaconis1988group,kondor2010ranking} can be expressed as the learning of permutations \cite{kondor2008group,jiao2015kendall}. Other examples are learning problems in coset spaces such as partial rankings, Q-sort data, error correcting codes and homogeneous spaces  \cite{diaconis1988group}. 
\begin{figure*}[t!]
	\centering
	\includegraphics[width=0.95\textwidth]{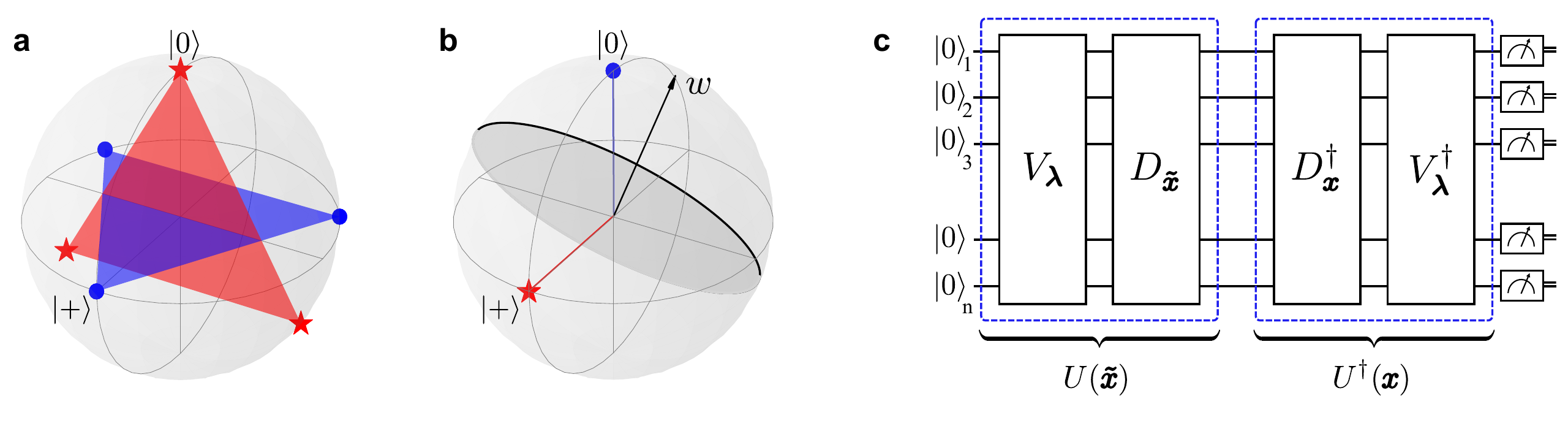}
	    \caption{\textbf{Labeling cosets}. (a), (b) Two covariant feature maps for a single-qubit example of the \emph{labeling cosets} learning problem introduced in the text. We take $S = \{\mathds{1}, A, A^2\}$ as subgroup of $G = SU(2)$, where $A = \exp(i (2\pi/3) X)$. Choosing two elements $\myVec{c}_+, \myVec{c}_- \in SU(2)$, with their representation $D_{\myVec{c}_+}=\mathds{1}$, $D_{\myVec{c}_-}=H$, we form two left-cosets: $C_+ = D_{\myVec{c}_+} S = S$ and $C_- = D_{\myVec{c}_-} S = \{H, HA, HA^2\}$. (a) Non-linearly-separable case with fiducial state $|\psi\rangle = |0\rangle$. Data points $\myVec{x} \in C_{\pm}$, for which $\myVec{x} = \myVec{c}_{\pm} \myVec{s}$, are mapped to the states $D_{\myVec{c}_+} D_{\myVec{s}} |0\rangle$ and $D_{\myVec{c}_-} D_{\myVec{s}} |0\rangle$, for all $\myVec{s} \in S$, which correspond to points on the Bloch sphere marked by three red stars and three blue dots, respectively. As a result, elements from different cosets live on orthogonal planes and cannot be linearly separated. (b) Separable case with fiducial state as a subgroup invariant state $|\psi\rangle = |+\rangle$. Elements from a given coset are mapped to a single state: $D_{\myVec{c}_+} D_{\myVec{s}} |+\rangle = D_{\myVec{c}_+}|+\rangle = |+\rangle$ or $D_{\myVec{c}_-} D_{\myVec{s}} |+\rangle = D_{\myVec{c}_-}|+\rangle = |0\rangle$ for all $\myVec{s} \in S$. A classifier needs only to distinguish between two points $|+\rangle$ and $|0\rangle$, which are linearly separable by an optimal hyperplane tilted at 45$^{\circ}$ from the $zy$ plane.(c) Quantum circuit for calculating matrix elements of a covariant quantum kernel. The feature map $U(\myVec{x})$ is defined in terms of a unitary representation $D_{\myVec{x}}$ for a group $G$ with $\myVec{x} \in G$ and a circuit $V_{\myVec{\lambda}}$ that prepares the fiducial state $|\psi_{\myVec{\lambda}}\rangle = V_{\myVec{\lambda}} |0^n\rangle$. The frequency of measuring all zeros in the computational basis is an estimate for the kernel matrix element $K_{\myVec{\lambda}}(\myVec{x}, \myVec{\tilde{x}})$}
    \label{fig:kernel}
\end{figure*}
We consider a general class of feature map circuits that we call {\it covariant feature maps} and that can be used for data space with a group structure. The corresponding quantum feature states are intimately related to covariant measurements \cite{holevo1979covariant}. The covariant feature map is defined relative to a unitary representation \cite{serre1977linear} $D_{\myVec{x}}$ for the group $G$ with $\myVec{x} \in G$, and a fiducial state $\ket{\psi} \in \mathbbm{C}^{2^n}$ on $n$ qubits as $\Phi(\myVec{x}) = D_{\myVec{x}} \proj{\psi} D_{\myVec{x}}^\dagger$. The {\it covariant quantum kernel} is then estimated as the fidelity [see Eq.~\eqref{eqn:kernel-definition}]
\bq\label{cov_kern}
K(\myVec{x},\tilde{\myVec{x}})  = |\bra{\psi} D^\dagger_{\myVec{x}} D_{\tilde{\myVec{x}}} \ket{\psi}|^2.
\eq
We assume the fiducial state $\ket{\psi} = V \ket{0^n}$ can be prepared by applying an efficient quantum circuit $V$. Likewise, it is important to also assume that the representation $D$ of $G$ can be implemented efficiently on a quantum computer. In this case, the QKE routine reduces to estimating the transition amplitude $K(\myVec{x}, \tilde{\myVec{x}}) = |\bra{0^n} V^\dagger D_{\myVec{x}}^\dagger D_{\tilde{\myVec{x}}}V\ket{0^n}|^2$, and the feature map circuit becomes $U(\myVec{x}) = D_{\myVec{x}} V$; cf. Fig.~\ref{fig:kernel}(c). The kernel as defined here is left-invariant under the group action. A right-invariant definition (c.f. Appendix B) is immediate. 

Covariant quantum kernels can lead to a provable separation between quantum and classical learners \cite{liu2020rigorous} for specific problems. The learning problem in \cite{liu2020rigorous} is a binary classification problem for data from the group $\mathbbm{Z}^*_p$ (integer multiplication modular $p$) and reduces to the discrete logarithm problem (DLOG) \cite{rosen2011elementary}. The kernel in \cite{liu2020rigorous} is a special case of the covariant quantum kernel introduced here, with the regular representation of $\mathbbm{Z}^*_p$ and a fiducial state that is the uniform superposition of group elements obtained from applications of the generator (c.f. Appendix B.1). In this example, it becomes apparent that different fiducial states can lead to vastly different kernels. While the aforementioned fiducial state produces a kernel that can lead to an efficient learning problem for the DLOG classification problem, a fiducial state that is given by one of computational basis states would lead to an identity kernel, matrix which is well-known to have extremely poor performance. This illustrates that the choice of $\ket{\psi}$ is essential for the performance of the quantum kernel. 
\begin{figure*}[t!]
	\centering
	\includegraphics[width=\textwidth]{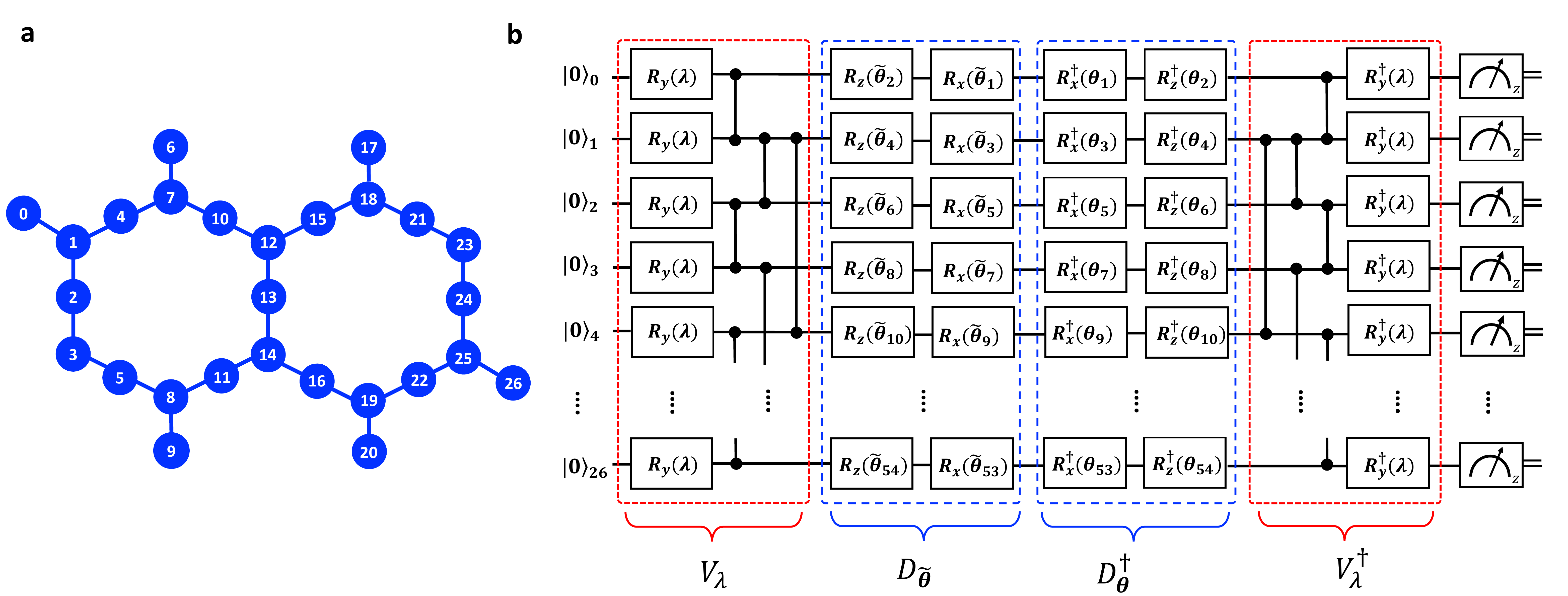}
	\caption{\label{Fig:Chip_Circuit} \textbf{Device layout and circuit mapping.}
	(a) The connectivity of the 27-qubit device \emph{ibmq\_kolkata}.
	(b) The quantum circuit used to evaluate the kernel matrix elements for the learning problem \emph{labeling cosets with errors}. Here, we define the single-qubit rotations as $R_{P}(\phi) = \exp(-i (\phi/2) P)$ for $P \in \{X,Y,Z\}$. We choose the representation of the group $G=SU(2)^{\otimes 27}$ to be $D_{\myVec{\theta}} = \bigotimes_{k=1}^{27} R_X(\theta_{2k-1}) R_Z(\theta_{2k})$. The fiducial state~\eqref{eqn:fiducial-state} is prepared by the circuit $V_\lambda$ for a single parameter $\lambda$. The entanglers $\mathrm{CZ}$, which are the controlled phase gates diag$(1,1,1,-1)$, match the connectivity of the quantum device in (a). The frequency of measuring all zeros in the computational basis is an estimate for a kernel matrix element.}
\end{figure*}
If sufficient structural knowledge about the problem is present a suitable fiducial state can be chosen \emph{a priori}. However, we also want a method to optimize the fiducial state subject to the available data, if no prior knowledge is available. The objective of this optimization will depend on the learning problem. We consider a binary classification problem with SVMs. For other types of kernel functions, objectives have been proposed \cite{lloyd2020quantum,otten2020quantum} that are motivated by quantum information theoretic insights. Here, we will follow a method commonly used in the classical literature referred to as {\it kernel alignment}  ~\cite{cristianini2001kernel,cortes2012algorithms}. 

To optimize the fiducial state, it is first generated by applying a variational quantum circuit $V_{\myVec{\lambda}}$ with  parameters $\myVec{\lambda} \in \Omega \subset \bR^q$ to the state $\ket{\psi_{\myVec{\lambda}}} = V_{\myVec{\lambda}} \ket{0^n}$. This will lead to parametrized quantum kernels $K_{\myVec{\lambda}}(\myVec{x},\myVec{\tilde{x}})$ with feature map circuit $D_\myVec{x}V_{\myVec{\lambda}}$ as depicted in Fig.~\ref{fig:kernel}(c). The parameters $\myVec{\lambda}$ are optimized with kernel alignment. The binary classifier associated with kernel $K_{\myVec{\lambda}}$  is given as a linear threshold function $f(\myVec{x}) = \mbox{sign}(\sum_{i=1}^m y_i \alpha_i K_{\myVec{\lambda}}(\tilde{\myVec{x}}_i,\myVec{x}))$ with model parameters $\{\alpha_i\}_{i=1\ldots m}$ for a training set of size $|\{\myVec{x}_i\}| = m$ and labels $y_ i = \pm 1$.  We use  a ``weighted'' version of the alignment  \cite{cortes2012algorithms,bullins2017not} to optimize the kernel parameters while the SVM is used to optimize the model parameters. This approach can be seen as optimizing the SVM upper bound on the generalization error directly (c.f. Appendix A). The cost function
\begin{equation}
\label{eqn:svm-objective}
	F(\myVec{\alpha}, \myVec{\lambda}) = \sum_{i=1}^m \alpha_i - \frac12 \sum_{i,j=1}^m \alpha_i \alpha_j y_i y_j K_{\myVec{\lambda}}(\myVec{x}_i, \myVec{x}_j),
\end{equation}
is related to an upper bound to the generalization error when maximized over $\myVec{\alpha}$. The weighted kernel alignment minimizes this upper bound with respect to $\myVec{\lambda}$. The procedure is expressed as the optimization, $\min_{\myVec{\lambda}} \max_{\myVec{\alpha}} F(\myVec{\alpha}, \myVec{\lambda})$,  subject to the constraints of the feasible set $0 \leq \alpha_i \leq C$, where $C$ is the box parameter, $\sum_i y_i \alpha_i = 0$, and $\myVec{\lambda} \in \Omega$. In the Supplementary Material (c.f. Appendix D.2), we present a stochastic algorithm for this optimization problem, which is an iterative algorithm with kernel matrices evaluated on a quantum processor and continuous parameters updated with classical optimization routines.

In the quantum experiment we want to benchmark both the accuracy of a learner with access to covariant quantum kernels, as well as the performance of the physical hardware. It is therefore important to have a data set that allows for zero classification error. We therefore construct an artificial benchmark data set. We introduce a learning problem  {\it labeling cosets with error} (LCE) that serves as an abstraction of common learning problems on coset spaces (c.f. Appendix C). Learning problems on coset spaces are frequently considered in the literature~\cite{diaconis1988group,kondor2008group}, for example when considering partial rankings \cite{kondor2010ranking} or for manifolds that arise as homogeneous spaces \cite{diaconis1988group}. In LCE we are given a group $G$ and subgroup $S < G$ and generate data from two left-cosets, $c_{\pm}S \subset G$ determined by representatives $c_{+},c_{-} \in G$. See Fig.~\ref{fig:kernel}(a) for a single-qubit example. Every datum taken from cosets is perturbed with a small error $\epsilon$ so that the data is not part of the coset any longer. After seeing sufficient data, the learner is asked to classify to which coset a previously unseen datum belongs.
\begin{figure*}[t!]
	\centering
	\includegraphics[width=0.9\textwidth]{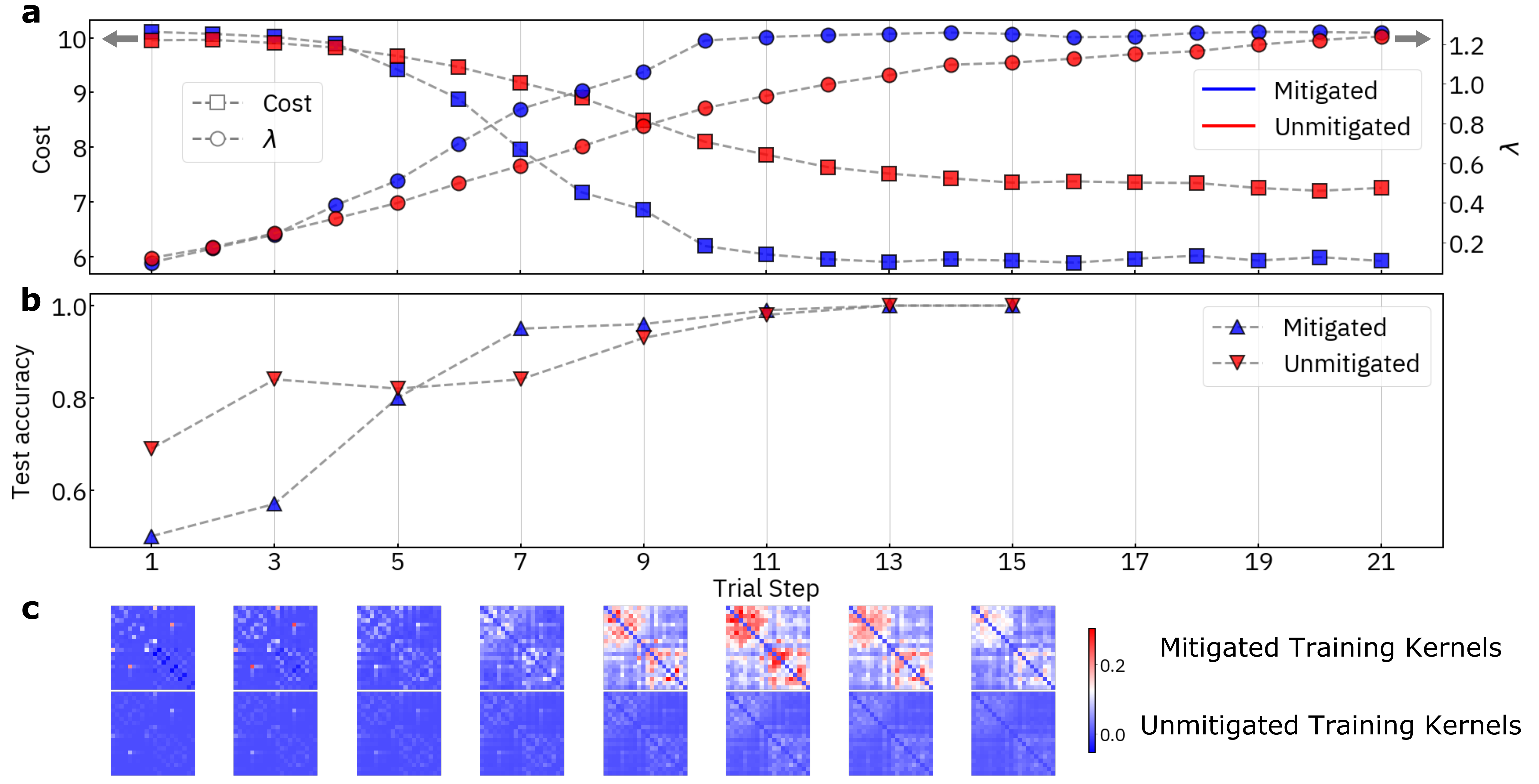}
	\caption{\label{FigExp}  \textbf{Kernel alignment.} (a) Kernel alignment cost function~\eqref{eqn:svm-objective} and fiducial state parameter $\lambda$ for each SPSA trial step, for the error mitigated (blue symbols) and unmitigated (red) experimental realizations. The ideal $\lambda$ is $\pi/2$. The training data set during this optimization contains 10 data points per label. (b) Accuracy tests results at odd trial steps. For these tests, a training set kernel is estimated with the $\lambda$ parameter at that step, from which a separating hyperplane is obtained and used to classify a test set consisting of 50 data points per label. The training set kernels used for each classification are shown in (c) for both the mitigated and the unmitigated cases. The identity matrix has been subtracted from these kernels for clarity.}
\end{figure*}
We implement LCE with a group $G$ motivated by our physical hardware. We implement our kernel alignment, training, and classification test experiments on a $n = 27$-qubit superconducting quantum processor with heavy-hexagon topology, c.f. Fig.~\ref{Fig:Chip_Circuit}(a)~\cite{chamberland2020topo, jurcevic2021demonstration}. We choose $G = \mathsf{SU}(2)^{\otimes n}$ with the natural representation $D$ of $\mathsf{SU}(2)$ for each qubit, and a Pauli-stabilizer group $S = \avr{s_1,\ldots,s_n}$ \cite{gottesman1997stabilizer} as subgroup of $G$. In particular, we choose the graph-stabilizer \cite{hein2006entanglement}
\be\label{stab-graph}
    S_{\mbox{graph}} = \avr{\{X_i \bigotimes_{k : (k,i) \in E} Z_k\}_{i \in V}}
\ee
associated with the coupling graph $(E,V)$ given by the device connectivity.
To generate the data we represent the rotations $D(\theta_1,\theta_2,\theta_3) \in  \mathsf{SU}(2)$ by their Euler angles $D(\theta_1,\theta_2,\theta_3) = \exp(-i (\theta_1/2) X) \exp(-i (\theta_2/2) Z) \exp(-i (\theta_3/2) X)$. For simplicity we set all $\theta_3 = 0$ and randomly draw two $\myVec{c}_{\pm} \in [-\pi/2,\pi/2]^{2n}$ and define each class by the cosets $C_{\pm} = \myVec{c}_{\pm} S_{\mbox{graph}}$. The rotations with the representative angles $\myVec{c}_{\pm}$ can be combined with the elements from the stabilizer group so that each datum in the coset is expressed as the rotation $D_{\myVec{\theta}} = \bigotimes_{k=1}^{n} D(\theta_{2k-1},\theta_{2k},0)$. We express every element in the coset in terms of the Euler angles $\myVec{\theta}$. To generate data for training and testing we uniformly sample elements from $C_{\pm}$ and perturb each set of Euler angles with a normal random error of variance $\epsilon = 0.01$. 

After having fixed the unitary representation, we need to choose a fiducial state family. A good a priori choice is a state that is invariant under the sub-group action $D_{\myVec{s}} \ket{\psi} = \ket{\psi}$ for all $\myVec{s} \in S$, illustrated in Fig.~\ref{fig:kernel}(b) in the form of a single-qubit example. For $S$ as in (\ref{stab-graph}) this state is given by a stabilizer graph state \cite{hein2006entanglement}. Our benchmark for kernel alignment asks to recover this graph state and a perfect classification accuracy from the optimization of the variational state family
\be
\label{eqn:fiducial-state}
    \ket{\psi_\lambda} = \prod_{(k,t) \in E} \mbox{CZ}_{k,t} \prod_{k \in V} R_{Y_k}(\lambda)\ket{0^n},
\ee
where $R_{Y}(\lambda) = \exp(-i (\lambda/2) Y)$ for a uniform $\lambda \in [0,2\pi]$. This state interpolates between a trivial initial state $\ket{\psi_0} = \ket{0^n}$ and a graph state stabilized by the subgroup. The resulting covariant quantum kernel, cf. Fig.~\ref{Fig:Chip_Circuit}(b), is given as $K_{\lambda} (\myVec{\theta} , \tilde{\myVec{\theta}}) = |\bra{\psi_\lambda} D^\dagger_{\myVec{\theta}} D_{\tilde{\myVec{\theta}}} \ket{\psi_\lambda}|^2$.

We run kernel alignment with and without error mitigation \cite{temme2017error,li2017efficient, kandala2019error}. The stochastic optimization algorithm for $\lambda$ uses Simultaneous Perturbation Stochastic Approximation (SPSA) in both cases. We cover 21 trial steps, starting from $\lambda = 0.1$ in both experimental realizations, using the same training set consisting of 10 data points per label. As we can see in Fig.~\ref{FigExp}(a), the cost, Eq.~\eqref{eqn:svm-objective}, flattens after around 11 SPSA trials both for the mitigated an unmitigated cases, with a lower cost for the mitigated experiments. The kernel parameter $\lambda$ approaches its ideal value of $\pi/2$ at a faster rate in the mitigated case. Once the training is completed, accuracy tests can be run on different sets of data points. We run these classification tests at each odd SPSA step for the mitigated and the unmitigated protocols, targeting a test set with 50 data points per label. As shown in Fig. ~\ref{FigExp}(b), the mitigated experiments reach high classification success percentages faster than the unmitigated version, although both approaches reach 100\% success after SPSA step 13. We show the training set Gram matrices at odd steps of SPSA for the mitigated and unmitigated approaches in Fig.~\ref{FigExp}(c). Each circuit in all these experiments is sampled 8,192 times.\\

{\it Conclusions:}  We have identified a promising class of learning problems that stand to profit from the use of quantum kernels. Learning problems with a group theoretic structure have important practical applications \cite{diaconis1988group,kondor2008group} and provide a high degree of structure that can be exploited in quantum algorithm design. The quantum kernel method makes it possible to turn a group theoretic learning problem in to a geometric question. Advanced classical kernels for group-data \cite{kondor2008group} are known to be computationally expensive to evaluate, which is why they are mostly approximated in terms of their lowest Fourier-weights \cite{kondor2010ranking,huang2009fourier}. A potential advantage of quantum kernels can be that such an approximation may not be necessary. The learning problem we have studied experimentally serves as an abstract representation of many group theoretic problems, i.e. classification of data that is close to cosets of a group.  As discussed in the manuscript, we find that an important degree of freedom for the covariant quantum kernels is the fiducial state. We expect that, depending on the problem, different choices have to be made. The data for the problem was generated artificially, so that we could ensure that the classification problem could be solved to arbitrary precision and we have a high fidelity benchmark of the experiment. The specific learning problem was constructed with groups that can be implemented naturally and efficiently on current quantum hardware with, to date, the largest circuit width for a quantum kernel. The considered circuit, although supported on a larger number of qubits, is still sufficiently shallow so that classical simulation methods can be applied \cite{bravyi2021classical}. These methods exploit the explicit two-dimensional structure, an advantage that disappears for deeper circuits or other coupling topologies.

The core element of the QKE routine reduces to the estimation of an expectation value. If the circuit remains within the coherence limit set by the device, error mitigation techniques \cite{temme2017error,li2017efficient, kandala2019error} can be applied. This has been demonstrated in the $27$ qubit experiment presented here. The experiment shows that even for a $27$ qubit experiment the kernel matrix can be significantly improved by using error mitigation techniques. Note that no readout error mitigation~\cite{bravyi2021mitigating} has been applied here, giving rise to the hope that, using threshold functions of the kernels, this approach is indeed robust against small experimental imperfections. For regression problems \cite{huang2020power} or other applications where the actual magnitude of the kernel is more important, error mitigation schemes may be increasingly important.\\

{\it Acknowledgements:} We thank Srinivasan Arunachalam and Vojtech Havlicek for helpful comments and discussions and Doug T. McClure and Neereja Sundaresan for technical assistance with the device. This work was supported by the IBM Research Frontiers Institute and K.T. acknowledges support from the ARO Grant W911NF-20-1-0014.

\onecolumngrid
\appendix

\section{Support vector machines}
Although kernels find a wide range of applications throughout the machine learning and statistics literature \cite{vapnik1996regression}, we focus in the manuscript specifically on the supervised learning learning problem of binary classification. More precisely we consider binary classification with support vector machines (SVM)s. We only review the most elementary components here, in particular in light of the fact that we use a quantum computer to evaluate the kernel entries in the otherwise classical algorithm.  In binary classification one considers two data sets (or a partition into two sets) as the training set $T = \{\myVec{x}_i,y_i\}_{i=1\ldots m} \subset {\cal X} \times \bZ_2$ for a data space ${\cal X}$ and the testing set $S \subset {\cal X} \times \bZ_2$, where typically $S \cap T = \emptyset$. Support vector machines for data that is not linearly separable \cite{vapnik2013nature} usually employ a feature mapping to map $\Phi : {\cal X} \rightarrow \bR^N$ from the original data space ${\cal X}$ into a higher dimensional Euclidean vector space $\bR^N$ with inner product $\langle \cdot, \cdot\rangle_{\bR^N}$, often referred to as feature space. The binary SVM classifier is a linear threshold function in this feature space 
\be\label{dec-primal}
    f(\myVec{z}) = \mbox{sign} \left( \langle  w, \Phi(\myVec{z}) \rangle_{\bR^N} + b\right),
\ee
where the vector $w \in \bR^N$ is referred to as the hyperplane normal vector that defines the linear threshold function and $b \in \bR$ is an offset referred to as a bias. The training of the classifier occurs when optimizing the upper bound of the generalization error \cite{shawe2002generalization} in terms of the soft margin of the binary classifier for the training set $T$. This is commonly expressed in terms of the following quadratic program
\begin{align}\label{SVM-primal}
    &\mbox{minimize} \;\;\; \frac{1}{2} \| w \|_2^2 + C \sum_{i=1}^m \xi_i \nonumber \\
    &\mbox{subject to:} \;\;\; y_i(\langle  w, \Phi(\myVec{x}_i) \rangle_{\bR^N}) + b \geq 1 - \xi_i   \nonumber \\ 
    & \;\;\;\;\;\;\;\;\;\;\;\;\;\;\;\;\;\;\;\; \xi_i \geq 0 \;\;\; \forall i =1 \ldots m
\end{align}
The box parameter $C \geq 0$ determines how strongly data points that are not linearly separable contribute to the cost function, where individual contributions for each data point are expressed in terms of the additional variable $\xi_i$. This is a convex problem and can be solved efficiently in the dimension $N$ of the feature space. However, it is frequently assumed that this dimension is large so that the problem, although convex, becomes intractable. To deal with this problem the kernel method is applied. The kernel function is just the inner product between two feature maps $K(\myVec{x}_i,\myVec{x}_j) = \langle \Phi(\myVec{x}_i), \Phi(\myVec{x}_j) \rangle_{\bR^N}$. The kernel appears for more general functions by means of the representer theorem~\cite{scholkopf2001generalized}. For support vector machines the kernel appears naturally when considering the dual optimization problem to Eq.~\eqref{SVM-primal}. The dual problem \cite{vapnik2013nature} for the primal problem with $m = |T|$ constraints is again a quadratic program in the Lagrange multipliers $\alpha_i$ for $i = 1 \ldots m$
\begin{align}\label{SVM-dual}
    &\mbox{maximize} \;\;\; \sum_{i=1}^m \alpha_i - \frac12 \sum_{i,j = 1}^m \alpha_i \alpha_j y_i y_j K(\myVec{x}_i,\myVec{x}_j) \nonumber \\
    &\mbox{subject to:} \;\;\; \sum_i y_i \alpha_i = 0   \nonumber \\ 
    & \;\;\;\;\;\;\;\;\;\;\;\;\;\;\;\;\;\;\;\; 0 \leq \alpha_i \leq C \;\;\; \forall i =1 \ldots m
\end{align}
The optimal dual variables $\alpha^*_i$ are related to the optimal primal normal $w^* = \sum_{i=1}^m y_i \alpha^*_i \Phi(\myVec{x}_i)$ so that the decision function (\ref{dec-primal}) can be expressed in terms of the kernel as well. For a new datum $\myVec{z} \notin T$ we write the decision function  
\be
\label{eqn:svm-decision-function}
    f(\myVec{z}) = \mbox{sign}\left(\sum_{i=1}^m y_i \alpha_i K(\myVec{x}_i,\myVec{z}) + b\right).
\ee
Both training of the SVM as in Eq.~\eqref{SVM-dual} as well as classification~\eqref{eqn:svm-decision-function} can be performed efficiently in arbitrary spatial dimensions $N$, even in infinite-dimensional Hilbert spaces, if the kernel function can be evaluated efficiently.

\subsection{Quantum kernels and quantum feature maps}
The motivation to introduce quantum kernels in \cite{havlivcek2019supervised} stems from the observation that many of the frequently considered quantum machine learning models, such as variational quantum neural networks \cite{mitarai2018quantum,farhi2018classification,schuld2020circuit} can be understood as simple kernel models. This means that if the data mapping in these models leads to states for which the fidelity can be estimated efficiently up to an additive sampling error on a classical computer, conventional support vector machine methods can be applied and it is not possible to obtain a quantum speedup. The main motivation is therefore to consider quantum feature maps that lead to quantum kernel functions that are classically hard to evaluate but can be approximated on a quantum computer. With such a quantum kernel function, the supervised learning algorithm is the direct application of the SVM algorithm as explained in the previous section. The only difference is the use of a quantum computer in estimating the kernel function. Following \cite{havlivcek2019supervised} this kernel function can be understood as the $\ket{0^n}$ to $\ket{0^n}$ transition probability of a particular unitary quantum circuit on $n$ qubits. The non-linear feature mapping for a datum $\myVec{x} \in {\cal X}$ occurs through the application of the datum dependent unitary circuit $U(\myVec{x})$ to a reference state $\ket{0^n}\bra{0^n}$. The resulting feature vector is the quantum state of density matrix 
\be
    \Phi(\myVec{x}) = U(\myVec{x})\proj{0^n}U ^\dagger(\myVec{x}).
\ee
When fitted with conventional trace inner product $\langle A, B \rangle_{tr} = \tr{A^\dagger B} $ the space of Hermitian matrices becomes a Euclidean vector space. The space that is obtained by the aforementioned construction is a $N = 4^n-1$ - dimensional Euclidean space that can be mapped directly on to $\bR^N$ when a suitable Hermitian matrix basis is chosen. One such example is the basis of Pauli matrices \cite{nielsen2010quantum}. The features are the components of the feature vectors, relative to this basis and reduce to the expectation values of the Hermitian matrices with respect to the  state $\Phi(\myVec{x})$. The kernel function is then the trace inner product between $K(\myVec{x},\myVec{z}) = \langle \Phi(\myVec{x}), \Phi(\myVec{z}) \rangle_{tr}$ and evaluates to the transition amplitude 
\be
    K(\myVec{x},\myVec{z}) = |\bra{0^n} U^\dagger(\myVec{x})U(\myVec{z}) \ket{0^n}|^2. 
\ee
This can be estimated up to an additive error by first preparing the state $\ket{0^n}$ applying the circuit $U^\dagger(\myVec{x})U(\myVec{z})$ and then counting the number of times the all-zero outcome $0^n$ is observed. Other ways of estimating this kernel are through the use of a SWAP test \cite{cincio2018learning,havlivcek2019supervised}. This method may become relevant when considering larger noisy devices.

While it may appear to be only a detail, it is actually important to understand the feature state as the density matrix $\Phi(\myVec{x})$. This ensures that the feature space is a Euclidean vector space and it determines the form of the algorithm to access the feature space \cite{havlivcek2019supervised}. Formally using only $\ket{\phi(x)} = U(\myVec{x})\ket{0^n}$ and the corresponding canonical inner product $\langle \phi(x)|\phi(z)\rangle$ \cite{schuld2019quantum} would lead to a kernel and classifiers that are sensitive to the global phase of the state. 

\section{Covariant quantum kernels}
In this manuscript we introduce a class of quantum kernels suited for learning tasks with data sets that can be seen as a subset of some group ${\cal X} \subseteq G$, where the group $G$ is determined by the learning problem at hand. The kernel is computed from a feature mapping that is inspired by covariant quantum measures \cite{holevo1979covariant}. To define the feature map in general, we need two components:

\begin{enumerate}
    \item{\bf Unitary representation:} We define a quantum circuit $D_\myVec{x} \in U(2^n)$ that corresponds to a unitary representation $D : {\cal X} \rightarrow U(2^n)$ \cite{serre1977linear,diaconis1988group} of the group $G$ that will act on $n$ qubits. A quantum circuit is assigned for every datum $\myVec{x} \in {\cal X}$ and we require that this quantum circuit be implemented efficiently on the quantum computer. 
    
    \item{\bf Fiducial state:} Furthermore, we require a reference state $\ket{\psi} \in \left(\bC^{2}\right)^{\otimes n}$ on which the representation can act. This state is referred to as a fiducial state and we assume that it can be prepared efficiently 
    by a quantum circuit $V$, so that  $\ket{\psi} = V \ket{0^n}$.
\end{enumerate}
Two examples will be presented in Secs.~\ref{sec:formal_sep} and \ref{sec:coset-learning}. It will become apparent that the choice of fiducial state is important to achieving good classification accuracy. The quantum feature mapping can be defined in two similar ways. One $\Phi^l$, which leads to a {\it left-invariant} quantum kernel, and another $\Phi^r$, which leads to a {\it right-invariant} kernel function as will be discussed shortly. 
\begin{enumerate}
\item{\it Left-invariant feature state:} 
\begin{equation}
    \Phi^l: \myVec{x} \rightarrow \Phi^l(\myVec{x}) = D_\myVec{x} \proj{\psi} D^\dagger_\myVec{x}
\end{equation}
\item{\it Right-invariant feature state:} 
\begin{equation}
    \Phi^r: \myVec{x} \rightarrow \Phi^r(\myVec{x}) = D^\dagger_\myVec{x} \proj{\psi} D_\myVec{x}
\end{equation}
\end{enumerate}
In keeping with the notation for quantum feature maps, we identify the quantum feature map circuit as $U(\myVec{x}) = D_{\myVec{x}}V$ and $U(\myVec{x}) = D^\dagger_{\myVec{x}}V$ respectively. This feature mapping can then be used to define the following kernel functions $K^l(\myVec{x},\myVec{z}) = \tr{\Phi^l(\myVec{x})\Phi^l(\myVec{z})}$ and $K^r(\myVec{x},\myVec{z}) = \tr{\Phi^r(\myVec{x})\Phi^r(\myVec{z})}$ which are given by
\begin{enumerate}
\item{\it Left-invariant kernel:} 
\begin{equation}
    K^l(\myVec{x},\myVec{z}) = |\bra{\psi} D^\dagger_{\myVec{x}} D_{\myVec{z}} \ket{\psi} |^2 = |\bra{0^n} V^\dagger D^\dagger_{\myVec{x}} D_{\myVec{z}} V \ket{0^n} |^2.
\end{equation}
\item{\it Right-invariant kernel:} 
\begin{equation}
    K^r(\myVec{x},\myVec{z}) = |\bra{\psi} D_{\myVec{x}} D^\dagger_{\myVec{z}} \ket{\psi} |^2 = |\bra{0^n} V^\dagger D_{\myVec{x}} D^\dagger_{\myVec{z}} V \ket{0^n} |^2.
\end{equation}
\end{enumerate}
The feature maps as well as the kernel functions are referred to as {\it right}- and {\it left-invariant} based on their invariance to group multiplication. Consider for any $\myVec{g} \in G$ the functions  $K^l(\myVec{g}\myVec{x},\myVec{g}\myVec{z}) = |\bra{\psi} D^\dagger_{\myVec{x}} D^\dagger_{\myVec{g}} D_{\myVec{g}} D_{\myVec{z}} \ket{\psi} |^2  = |\bra{\psi} D^\dagger_{\myVec{x}} D_{\myVec{z}} \ket{\psi} |^2 = K^l(\myVec{x},\myVec{z})$ and $K^r(\myVec{x} \myVec{g},\myVec{z}\myVec{g}) = |\bra{\psi} D_{\myVec{x}} D_{\myVec{g}} D^\dagger_{\myVec{g}} D^\dagger_{\myVec{z}} \ket{\psi} |^2  = |\bra{\psi} D_{\myVec{x}} D^\dagger_{\myVec{z}} \ket{\psi} |^2 = K^r(\myVec{x},\myVec{z})$ respectively.  This means that when $\myVec{1} \in G$ is the identity element, the kernel functions can always be expressed as  $K^r(\myVec{x},\myVec{z}) = K^r(\myVec{x}\myVec{z}^{-1}, \myVec{1})$ and $K^l(\myVec{x},\myVec{z}) = K^l(\myVec{z}^{-1}\myVec{x}, \myVec{1})$ respectively. Hence the kernel function is in either case determined by a single function $f^{r/l}(\myVec{g}) =   K^{r/l}(\myVec{g}, \myVec{1})$, and can be recovered from setting either $\myVec{g} = \myVec{z}^{-1}\myVec{x}$ or $\myVec{g} = \myVec{x}\myVec{z}^{-1}$. \\ 

For notational simplicity and to be consistent with the main body of the paper, let us only focus on the {\it left-invariant} kernel functions $f^{l}(\myVec{g}) = K^{l}(\myVec{g}, \myVec{1})$. We will from now on simply refer to this function by $l(\myVec{g})$. Recall that a unitary representation $D$ of a group $G$ can always be decomposed in to a direct sum of a set ${\cal J}_{D}$ of  irreducible representations $D^J$ by a basis change \cite{serre1977linear,diaconis1988group}
\be
D_\myVec{g} \simeq \bigoplus_{J \in {\cal J}_{D}} D^J_\myVec{g} \;.
\ee
Given such a decomposition into irreducible representations, general functions on the group, i.e. $f : G \rightarrow \bR$, can be expressed in terms of the non-Abelian Fourier transform via
\be
\hat{f}(J) =  \sum_{\myVec{g} \in G} f(\myVec{g}) D^J_{\myVec{g}}.
\ee
Recall that $\hat{f}(J) \in M_{\mbox{d}_J \times \mbox{d}_J}(\bC)$ is a matrix-valued function, where the matrix dimension is determined by the dimension of the irreducible representation $\mbox{d}_J$. The inverse non-Abelian Fourier transform is given by
\be\label{supp-eq:FTinversion}
f(\myVec{g}) = \frac{1}{|G|} \sum_{J} \mbox{d}_J \tr{\hat{f}(J){D^J}_{\myVec{g}^{-1}}}.
\ee
Hence, we will be able to represent the covariant quantum kernel $l(\myVec{x})$ in terms of its Fourier coefficients $\hat{l}(J)$. These Fourier coefficients are determined by the fiducial state $\ket{\psi}$ and the choice of representation $D_{\myVec{x}}$. One can easily verify that 
\be
    K^l(\myVec{x},\myVec{1}) = |\bra{\psi} D^\dagger_{\myVec{x}} \ket{\psi}|^2 = \bra{\psi,\overline{\psi}} D_{\myVec{x}^{-1}} \otimes \overline{D}_{\myVec{x}^{-1}}
    \ket{\psi,\overline{\psi}},
\ee
where bar indicates complex conjugate. Note that the general tensor product representation $D_{\myVec{x}} \otimes \overline{D}_{\myVec{x}}$ between the original representation and its conjugate can also be decomposed in to irreducible representations and a similar block decomposition exists
\be
    D_{\myVec{x}} \otimes \overline{D}_{\myVec{x}} \simeq  \bigoplus_{J \in {\cal J}_{D\overline{D}}} D^J_{\myVec{x}}.
\ee
This can be exploited to evaluate the non-Abelian Fourier transform of the quantum kernel as 
\be\label{Sup-eqn:fourier-coeff}
    \hat{l}(J) = \frac{|G|}{d_J} \Pi_J \proj{\psi, \overline{\psi}} \Pi_J,
\ee
where the projector $\Pi_J$ is to be read as the projector on the subspace spanned by all the irreducible representations labeled by $J$. This projector is assumed to act on the space of the representation $D_{\myVec{x}} \otimes \overline{D}_{\myVec{x}} $, which is the originally doubled Hilbert space $\mathbbm{C}^{2^{2n}}$ of $n$ qubits and can be constructed by virtue of the characters $\chi_J(g)$ of the irreducible representation $J$. That is, the projector is computed as 
\be
    \Pi_J = \frac{d_J}{|G|}\sum_{g \in G} \overline{\chi}_{J}(g) \; D_g \otimes \overline{D}_g.
\ee
By inserting the Fourier coefficients, c.f. Eq.~(\ref{Sup-eqn:fourier-coeff}) into the inversion formula Eq.~(\ref{supp-eq:FTinversion}), we readily recover the full kernel since we have that $D_{\myVec{x}} \otimes \overline{D}_{\myVec{x}} = \sum_J \Pi_J D^J_{_{\myVec{x}}} \Pi_J$. Note that the Fourier coefficients are given by rank-one projectors induced by the fiducial state $\ket{\psi}$ restricted to the subspace of the irreducible representation $J$.

\subsection{Example with formal separation}\label{sec:formal_sep}
We review an example learning problem for this kernel family introduced in \cite{liu2020rigorous} to establish a formal separation between learning with quantum kernels and all other classical learners without access to quantum resources. In this example the group is given by $\mathbbm{Z}^*_p = \{1,2,\ldots,p-1\}$, i.e. the integers with multiplication modulo $p$. We consider a fixed generator $g \in \mathbbm{Z}^*_p$ that generates the full group through modular exponentiation.  Given the generator $g$ we will always be able to write any $\myVec{x}  = g^{\myVec{v}}$ with ${\myVec{v}} \in \{0,1,\ldots,p-1\}$. The inverse of this mapping, i.e. given $\myVec{x}$ compute $\myVec{v} = \mbox{DLOG}_{g}(\myVec{x})$ is referred to as the discrete logarithm (DLOG). It is generally assumed that the computation of the discrete logarithm is a computationally hard task for classical computers \cite{rosen2011elementary}. That is, any classical algorithm is assumed to scale super-polynomially in $n = \lceil \mbox{log}_2(p) \rceil$. The learning problem constructed in \cite{liu2020rigorous} provides a formal separation between classical and quantum learners with access to quantum kernel functions relative to the hardness assumption of the discrete logarithm problem.\\

The learning problem is given as follows: To construct a ground truth distribution and labeling rule, draw an  element ${\myVec{s}} \in \mathbbm{Z}^*_p$ uniformly at random. Note, there are now ${\cal O}(2^n)$ different labeling rules that can be learned. The data to be classified is drawn uniformly at random from $\mathbbm{Z}^*_p \subset \{0,1\}^n$. The labeling function then assigns the labels according to the following rule: every sampled element from the training set $\myVec{x} \in T \subset \mathbbm{Z}^*_p$ is assigned a label $y$ according to
\bq
y=\begin{cases}+1,&\text{if }\mbox{DLOG}_{g}( {\myVec{x}})\in[{\myVec{s}},{\myVec{s}}+\frac{p-3}{2}],\\ -1,
                  &\text{else.}\end{cases}
\eq
Here, addition is also taken modulo $p$. The labeled bit-strings can be generated efficiently classically and are handed to the learner as training and testing sets. It was shown \cite{liu2020rigorous} that no classical classifier can assign the correct labels of this decision rule with a probability more than $\half + \mbox{poly}(n)^{-1}$, relative to the hardness of the discrete logarithm problem by utilizing a result by Blum and Micali \cite{blum1984generate}. Conversely it was also proven that near-perfect classification accuracy could be obtained with polynomial effort when the appropriate quantum kernel function was used. \\

The feature states and kernel for which the quantum advantage has been proven are precisely in the form of the covariant feature map and quantum kernel as discussed in this paper. These feature states are constructed as follows: Let $D_{\myVec{x}}$ denote the regular representation of $\mathbbm{Z}^*_p \subset \{0,1\}^n$ defined on $n$ qubits that acts on a computational basis state $\ket{\myVec{z}} \in \mathbbm{C}^{2^n}$ through
\be
D_{\myVec{x}} \ket{\myVec{z}} = \ket{{\myVec{x}} \circ {\myVec{z}} \mbox{ mod } p}.
\ee
This group action is implemented by a circuit for modular multiplication as, for example, provided in \cite{markov2012constant}. We then need to define the fiducial state $\ket{\psi}$. For a generator $g$ we define the following subset state on $\{0,1\}^k \subset \mathbbm{Z}^*_{p}$ as the fiducial state
\be
\ket{\psi} = \frac{1}{2^{\frac{k}{2}}}\sum_{{\myVec{v}} \in \{0,1\}^k} \ket{g^{\myVec{v}}}.
\ee 
The feature state $\Phi({\myVec{x}}) = \proj{\phi({\myVec{x}})}$ used in \cite{liu2020rigorous} is then obtained as a covariant feature state by writing $\ket{\phi({\myVec{x}})} = D_{\myVec{x}}\ket{\psi} = 2^{-\frac{k}{2}}\sum_{{\myVec{v}} \in \{0,1\}^k} \ket{{\myVec{x}} g^{\myVec{v}}}$. The kernel is constructed in the canonical form, c.f. Eq.~(\ref{cov_kern}).\\

\subsubsection{The importance of the fiducial state choice} 
This example highlights the relevance of choosing a good reference state $\ket{\psi}$. To contrast the classification behaviour consider the scenario where one was to modify the kernel and select the computational basis state $\ket{\psi} = \ket{0^n}$ as the fiducial reference state. In this scenario the action of the representation is simply $D_{\myVec{x}} \ket{\psi} = \ket{\myVec{x}}$ and the kernel matrix
reduces to 
\be
K({\myVec{x}},\tilde{\myVec{x}}) = \delta_{{\myVec{x}},\tilde{\myVec{x}}}.
\ee
That is, the kernel becomes the identity matrix. The corresponding distance measure is simply the point-metric that agrees if two data points are identical and is zero if they are not. Apart from the fact that this kernel is trivial to implement classically, it is well known that it will lead to a very poor classifier \cite{vapnik2013nature}. \\

This example stresses the importance of the choice of $\ket{\psi}$ and demonstrates that an advantage can only be obtained when a suitable state is chosen. Moreover, the example highlights that the choice of $\ket{\psi}$ may strongly depend on the particular form of the learning problem and the ground truth distribution. This poses a challenge we seek to address in the next section. 
 
\section{The learning problem}\label{sec:coset-learning}
Learning problems on data with group structure have a long tradition in statistics \cite{diaconis1988group,kondor2008group,kondor2010ranking}. In this paper we benchmark the accuracy of the quantum kernel based binary classifier in a quantum computing experiment. 
The learning problem and data set we construct for the benchmark correspond to the generic classification problem on homogeneous spaces, c.f. \cite{diaconis1988group}, Chapter 6. The data problem is chosen to ensure that we can reach zero error. We refer to the particular learning problem considered in the manuscript as labeling cosets with error.\\

{\it labeling cosets with error:} Let us define the general problem for an arbitrary group $G$ and the homogeneous space induced by some subgroup $S$.  We can define two distinct left-cosets for two group elements ${\myVec{c}}_+,{\myVec{c}}_- \in G$ by
\begin{align}
C_+ &= {\myVec{c}}_+ S \;\;\;\;\;\; \mbox{and} \;\;\;\;\;\;  C_- = {\myVec{c}}_- S.
\end{align}
The group elements ${\myVec{c}}_{\pm}$ can be drawn at random, for example by the Haar measure of $G$, to generate different instances of the learning problem. Once two cosets have been determined, the ground truth for the binary classification is as follows: We need two distributions $Q^{\epsilon}_{\pm} : G \rightarrow \bR^+_0$ that have most of their mass on their respective coset $C_{\pm}$ but can deviate from the coset by a perturbation measured with respect to a small parameter $\epsilon$. To assign a meaning to this deviation it becomes necessary to establish metrics on the group space \cite{diaconis1988group}. We will be working with the matrix norm approach for faithful representations, i.e. $D_{\myVec{x}} \neq  D_{\myVec{y}}$ when $\myVec{x} \neq  \myVec{y}$, and a unitarily invariant norm $\| \cdot \|$ on the defining space of $D$. The representation dependent distance is
\begin{equation}
    d_D(\myVec{x}, \myVec{y}) = \| D_{\myVec{x}} - D_{\myVec{y}} \|
\end{equation}
We ask that the ground truth distribution $Q^{\epsilon}_{\pm}(\myVec{x})$, has a constant amount of its mass within an $\epsilon$ window $\min_{\myVec{g} \in C_{\pm}}d_D(\myVec{g}, \myVec{x}) \leq \epsilon$ around each coset.\\ 

We will generate a set $T$ of $|T| = m$ data samples for training and testing from such a ground truth distribution by the following steps. For each of the $i=1 \ldots m$ pick a label $y_i = \pm 1$ and chose the corresponding group element $\myVec{q}_i \in C_\pm$ uniformly at random. Then, we randomly choose a perturbation  $\myVec{e}_i \in G$ that  is close to the identity element $\myVec{1}$ so that it satisfies $ d_D(\myVec{e}_i, \myVec{1}) \leq \epsilon$. The distribution of the random perturbation will depend on the group $G$ and the representation $D$. 
Each datum $\myVec{x}_i$ is then generated from  $D_{\myVec{x}_i} = D_{\myVec{e}_i}D_{\myVec{q}_i}$. This datum will be added to the set $T$ so that the final set we use for training and testing is then $T = \{(\myVec{x}_1,y_1),\ldots,(\myVec{x}_m,y_m)\}$. \\

The addition of the perturbation makes this problem different from a purely group-theoretic question. In the error-free limit $\epsilon = 0$ the question amounts to asking whether a given element $\myVec{x}_i$ belongs to a particular coset. Such group theoretic problems are known to be classically hard and have been investigated in the quantum setting to prove formal separations. A particularly relevant example is the subgroup non-membership problem \cite{watrous2000succinct}. However, once we add a random perturbation that puts the labeled elements outside of the coset, the problem is no longer purely group theoretic on its own. The problem becomes a geometric learning problem in a Euclidean feature space where the mapping is motivated by group theoretic considerations. 

A particularly natural fit for the quantum kernel and fiducial state can be motivated from the error-free setting. The error-free setting motivates a natural candidate for the reference state $\ket{\psi}$. If this state can be prepared efficiently, we would want it to remain invariant under the action of the subgroup $S$: 
\begin{equation}
    D_{\myVec{s}} \ket{\psi} = \ket{\psi} \;\; \forall \; \myVec{s} \in S.
\end{equation}
For such a state, the data samples are mapped to a unique representing state for each coset, since  all elements $\myVec{x} \in C_{\pm}$ are of the form $\myVec{x} = \myVec{c}_{\pm}\myVec{s}$ with $\myVec{s} \in S$. We therefore have that 
\be
\label{eqn:subgroup-invariant}
 D_{\myVec{c}_{\pm}\myVec{s}}\ket{\psi} = D_{\myVec{c}_{\pm}} D_{\myVec{s}} \ket{\psi} = D_{\myVec{c}_{\pm}} \ket{\psi}.
\ee
In this case, the classifier effectively only needs to distinguish between the two states $D_{\myVec{c}_{\pm}}\proj{\psi} D_{\myVec{c}_{\pm}}^\dagger$. This situation changes, however, when we learn cosets with error. The added error is expected to act as a perturbation on each of the two states. If the perturbation is small, e.g. $\epsilon \ll 1$, one would expect that the states will still be classified correctly. Such a covariant quantum feature map turns the initially group theoretic question into a geometric question in quantum feature space and is able to address the perturbations. \\

The data set for the accuracy benchmark is motivated by the considerations above. In particular, recall to perform the benchmark of the classifier with covariant quantum kernel we need to ensure that we have to work with data that in principle permits an arbitrarily small classification error. For the experiment described in the main section of the paper we choose a natural and easy to implement group. We focus on the group of single-qubit rotations for $n$ qubits, so that $G = SU(2)^{\otimes n}$. A simple subgroup of this group is the Pauli group on $n$ qubits and, by extension, any stabilizer group \cite{schlingemann2001quantum,hein2004multiparty}. We choose the graph-stabilizer of the heavy hex lattice as said subgroup $S$. This stabilizer group fixes a graph state that follows the chip topology as its invariant state \cite{gottesman1997stabilizer}.

\section{Quantum kernel alignment algorithm}
\subsection{Interpretation of kernel alignment}
We use the kernel alignment procedure to chose the best fiducial state in the kernel function from a family of states $\ket{\psi_\myVec{\lambda}} = V_{\myVec{\lambda}}\ket{0^n}$. Here, we optimize over different variational quantum circuits $V_{\myVec{\lambda}}$ parametrized by some $V_{\myVec{\lambda}} \in \Omega$. The resulting kernel is then $K_{\myVec{\lambda}}(\myVec{x}, \myVec{z}) = |\bra{0^n}V^\dagger_{\myVec{\lambda}}D^\dagger_{\myVec{x}}D_{\myVec{z}} V_{\myVec{\lambda}}\ket{0^n}|^2$.
The kernel alignment procedure for obtaining the optimal hyperplane for data classification solves the problem 
\begin{equation}
\label{eqn:minmaxgame}
    \min_{\myVec{\lambda}} \max_{\myVec{\alpha}} F(\myVec{\alpha}, \myVec{\lambda}),
\end{equation}
subject to the usual constraints for $\myVec{\alpha}$ given in Eq.~\eqref{SVM-dual}. The objective we optimize in both the kernel parameters $\myVec{\lambda}$ and the Lagrange multipliers $\myVec{\alpha}$ is
\begin{equation}
\label{eqn:svm-objective-appendix}
	F(\myVec{\alpha}, \myVec{\lambda}) = \sum_i \alpha_i - \frac12 \sum_{i,j} \alpha_i \alpha_j y_i y_j K_{\myVec{\lambda}}(\myVec{x}_i, \myVec{x}_j).
\end{equation}
This min-max problem has an interesting interpretation as choosing a kernel among all the available $K_{\myVec{\lambda}}$ that minimizes the SVM bound on the classification error. The classification error $\mathbbm{P}(y \neq f(\myVec{z}))$ of $f$ is the probability that the classifier fails to predict the correct label $y$ subject to $(\myVec{z},y)$ being drawn from the true data distribution. This error was upper bounded by Shawe-Taylor and Cristianini~\cite{shawe2002generalization} in terms of the fat-shattering dimension for linear threshold functions. The fat-shattering bound is data dependent and encodes the generalization error of the classifier. The bound is, up to polylogarithmic factors, given by the primal cost function \cite{vapnik2013nature} of the SVM classifier.
The optimization of the primal SVM cost function or its Wolfe dual in kernel space, cf. Eq.~\eqref{SVM-dual}, considered here can be interpreted as minimizing this upper bound over all admissible threshold functions. For all $\myVec{\lambda}$ the maximum $F^*(\myVec{\lambda}) = \max_{\myVec{\alpha}} F(\myVec{\alpha}, \myVec{\lambda})$ yields the upper bound to the classification error:
\begin{equation}
    \mathbbm{P}(y \neq f(\myVec{z})) \leq \tilde{\cal O}(m^{-1} F^*(\myVec{\lambda})).    
\end{equation}
The notation $\tilde{\cal O}$ suppresses polylogarithmic factors. The optimization problem in Eq.~\eqref{eqn:minmaxgame} thus optimizes the parameters in  $K_{\myVec{\lambda}}$ to yield the smallest upper bound on the classification error. The kernel alignment procedure for fiducial states $\ket{\psi_{\myVec{\lambda}}}$ can be interpreted as the search for the fiducial state that gives rise to the best bound on the data-dependent generalization as measured in terms of the fat-shattering dimension for a linear threshold function with the given kernel family.

\subsection{Stochastic algorithm}
Our approach to solving the optimization problem with a dataset $T$ as stated in Eq.~(\ref{eqn:minmaxgame}) is described in Algorithm~\ref{algo:quantum-kernel-alignment}. This iterative classical-quantum algorithm evaluates kernel matrices on a quantum processor and updates the parameters $\myVec{\lambda}$ and $\myVec{\alpha}$ with a classical optimizer. At each iteration, kernel matrices are computed using the QKE routine~\cite{havlivcek2019supervised} on quantum circuits parametrized by kernel parameters $\myVec{\lambda}$. The optimal values $\myVec{\lambda}^{*}$ and $\myVec{\alpha}^{*}$ obtained can then be used in a standard SVM program to predict labels for a test set. 

The parametrization of quantum circuits for the kernel matrices leads, in general, to a highly non-convex objective function~\eqref{eqn:svm-objective-appendix} in the kernel parameters. We use simultaneous perturbation stochastic approximation (SPSA)~\cite{spall1992multivariate} for the minimization over the kernel parameters $\myVec{\lambda}$ in~\eqref{eqn:minmaxgame}. This method approximates the gradient of~\eqref{eqn:svm-objective-appendix} with respect to $\myVec{\lambda}$ using only two objective function evaluations $F(\myVec{\alpha}_\pm, \myVec{\lambda}_\pm)$ independent of the dimension of $\myVec{\lambda}$ and is suitable when measurements of the objective function are subject to stochastic fluctuations, which is the case for kernels evaluated on noisy quantum hardware. For the concave optimization over $\myVec{\alpha}$ in~\eqref{eqn:minmaxgame}, we use a standard classical solver \texttt{CVXOPT}~\cite{cvxopt2021}, which yields a unique solution for $\myVec{\alpha}$. 

\onecolumngrid

\begin{algorithm}[H]
\centering
\caption{\label{algo:quantum-kernel-alignment} Quantum Kernel Alignment: learning the maximum-margin kernel}
\begin{algorithmic}[1]
\State  {\bf Input} Training set $T = \{ \myVec{x}_i\in \bR^n\}_{i=1}^m$ with labels $y \in \{-1, 1\}^m$, quadratic program solver \texttt{qp} (e.g., \texttt{CVXOPT}).
\State  {\bf Parameters} Number of measurement shots $R$, box constraint $C>0$ (the SVM regularization parameter), initial kernel parameters $\myVec{\lambda}_0 \in \bR^q$, and SPSA steps $P$.
\State  Calibrate the quantum hardware to generate short depth quantum kernel circuits.
\State  Set initial values of the kernel parameters $\myVec{\lambda} = \myVec{\lambda}_0$ for the quantum kernel circuits.
	\For{\texttt{$i=0$ to $P$ }}
		\State Generate random vector $\Delta \in \{-1,1\}^q$.
		\State Evaluate $\myVec{\lambda}_{\pm,i} = \myVec{\lambda}_i \pm c_i \, \Delta$, where $c_i = c / (i+1)^\gamma$ for constants $c$, $\gamma$.
		\State Evaluate the kernel matrices $K_\pm = K(\myVec{\lambda}_{\pm,i}, T)$ on quantum device with $R$ measurement shots per~\cite{havlivcek2019supervised}.
        \State Maximize the SVM objective~\eqref{eqn:svm-objective-appendix} over $\myVec{\alpha}$ via \texttt{qp} solver $F(\myVec{\alpha}_{\pm,i}, \myVec{\lambda}_{\pm,i}) \leftarrow \texttt{qp}(K_\pm, y, C)$ subject to $0\le \myVec{\alpha}_\pm \le C$, $y^T \myVec{\alpha}_\pm = 0$.
		\State Update $\myVec{\lambda}_{i+1} \leftarrow \myVec{\lambda}_i - \frac{a_i}{2 c_i} \left[F(\myVec{\alpha}_{+,i}, \myVec{\lambda}_{+,i}) - F(\myVec{\alpha}_{-,i}, \myVec{\lambda}_{-,i}) \right]$ via SPSA, where $a_i = a / (i + 1 + A)^\sigma$ for constants $A$, $\sigma$.
	\EndFor
\State {\Return the final kernel parameters $\myVec{\lambda}^*$.}
\State Evaluate aligned kernel matrix $K(\myVec{\lambda}^*, T)$ on the quantum device with $R$ measurement shots per~\cite{havlivcek2019supervised}.
\end{algorithmic}
\end{algorithm}

While Algorithm~\ref{algo:quantum-kernel-alignment} is based on a simple implementation SPSA, other, more sophisticated optimizers can be used~\cite{rafique2018non,thekumparampil2019efficient,scholkopf2002sampling}. However, this choice is suitable to demonstrate the ideas we've presented here on an instance of the learning problem \emph{labelling cosets with errors}. The structural insights we have on samples drawn from the cosets are sufficient to inform a good choice of the parameterized fiducial state. As shown in Fig.~\ref{FigExp} from the main text, the parameter converges towards the expected value and the model reaches 100\% test accuracy on a 27-qubit problem instance on noisy hardware. In general, it important to have some information about the structure of the learning problem. Such insight can be used to inform the choice of parameterized fiducial state and increase the utility of QKA in practice.

\subsection{Additional considerations}
Here, we've considered weighted kernel alignment in our algorithm for learning kernels using the given data. A similar algorithm can be envisioned by replacing the weighted kernel alignment with alternatives like the unweighted kernel alignment \cite{cristianini2001kernel} and the centered kernel alignment \cite{cortes2012algorithms,Kim2006kernels, Gretton2005kernels}. Such a form of kernel alignment has recently been implemented for quantum kernels by \cite{hubregtsen2021training}. Optimizing unweighted kernel alignment, $\hat{A}$, between the ideal kernel and the desired kernel has a simple interpretation, which we can see from its definition:
\begin{equation}
\hat{A} \propto \sum_{i,j \in T} K(x_{i},x_{j})y_{i}y_{j} = \sum_{\substack{i \in \{1,2,..,m\} \\ j\in \{j|y_{i}=y_{j},i\neq j\}}}
K(x_{i},x_{j})-\sum_{\substack{i \in \{1,2,..,m\} \\ j\in \{j|y_{i}\neq y_{j}\}}}K(x_{i},x_{j})
\end{equation}
where $m$ is the number of training data points in set $T$. Optimizing the unweighted kernel alignment implies finding kernels that maximize the intra-class overlap and minimize the inter-class overlap for all points included in the training set $T$. Whereas, weighted kernel alignment Eq.~\eqref{eqn:minmaxgame} aims at finding a kernel using the given data that also maximizes the gap margin of the SVM classifier. This means that only the training points that are support vectors contribute towards learning the kernel. 

Previous studies suggest that kernel alignment based on kernels that have been normalized and centered in the feature space provide a more useful metric than uncentered kernel alignment \cite{cortes2012algorithms}.  As defined in \cite{cortes2012algorithms}, for a training set with $m$ data points, a centered kernel $K_{c}$ can be obtained from uncentered kernel $K$ via:
\begin{equation}
    K_{c} = \Big[ \boldsymbol{I} - \frac{\boldsymbol{1}\boldsymbol{1}^{T}}{m}\Big]K\Big[ \boldsymbol{I} - \frac{\boldsymbol{1}\boldsymbol{1}^{T}}{m}\Big]
\end{equation}
Centered kernel alignment then is defined exactly as the uncentered kernel alignment, using centered kernels instead. In general, the choice of optimizing centered kernel alignment leads to different alignment values when compared to uncentered kernel alignment. 

In the light of these choices, further studies are required to assess the performance of weighted and unweighted kernel alignment procedures with or without centered kernels. In addition, geometric differences between kernels and many other techniques developed for learning kernels from data in the classical machine literature may be used \cite{huang2020power, Srebro2006kernels, Jebara2006kernels, Kim2006kernels}. A natural follow up question would be a more exhaustive performance analysis of different kernel learning techniques applied in previous scenarios in the setting of covariant quantum kernels.
 
The algorithm outlined in this paper, involves optimization of free parameters in a parameterized quantum circuit. Variational methods that optimize over parameterized quantum circuits tend to suffer from the barren plateau problem \cite{McClean2018barren,Cerezo2020barren,Marrero2021barren} under certain conditions. The problem of barren plateaus is known to worsen when the cost function of the optimization problem relies on measurement of global, as opposed to local, observables~\cite{Cerezo2020barren}. We expect the barren plateau problem to be present in the implementation of this algorithm as the system size increases if the fiducial state is not chosen carefully. This can potentially be prevented by choosing a fiducial state that is well motivated with respect to the structure of the circuit, as well as the initial values of the parameters to be optimized.

\section{Experimental data}
\subsection{Device characterization}
The device (\textit{ibmq\_kolkata}, with the same topology as the device in Ref.~\cite{jurcevic2021demonstration}), consists of 27 fixed-frequency transmons with fundamental transition frequencies near 5 GHz and anharmonicities around -340 MHz, coupled by co-planar waveguide bus resonators on a top chip die, bump-bonded to a bottom die for readout and signal delivery. Single qubit gates are implemented via microwave pulses with Gaussian envelope with variance $\sigma$ equal to a quarter of the total pulse length and  with DRAG correction~\cite{motzoi2009simple} and two-qubit gates use the cross-resonance interaction ~\cite{Chow2011} with target rotary pulses ~\cite{Sundaresan2020}. The experiments described in this work take place over four consecutive days, and use a unique calibration for all the gates and readouts. The median qubit lifetime $T_1$ for the entire device is 132 $\mu$s, the median coherence time $T_2^{\mathrm{echo}}$ is 148 $\mu$s, and the median readout error (defined as $[P(0|1) + P(1|0)]/2$, where $P(i|j)$ represents the probability of measuring $i$ when the state is $j$) across all 27 qubits is 9.6e-3. During idle times of length $T_{\mathrm{idle}}$ for any qubit in our system we use a dynamical decoupling protocol whenever that idle time is longer than twice our single qubit gate. This dynamical decoupling consists of the following sequence: $\tau/2 - X_p - \tau - X_m - \tau/2$ with $\tau = (T_{\mathrm{idle}} - 2T_{X_{p/m}})/2$.
We exploit error mitigation techniques~\cite{temme2017error,li2017efficient, kandala2019error, kim2021errormit} in our experiments, using stretches $c=1$ and $c=1.3$, from which we apply a first-order zero-noise extrapolation to estimate the kernel entries $K_{\lambda} (\myVec{\theta} , \tilde{\myVec{\theta}})$. For stretch $c=1$ ($c=1.3$) the single qubit gates are 35.6 (46.2) ns long and the median CNOT length across the device is 476.4 (618.6) ns, with standard deviation of 134 (174) ns. The cross-resonance pulses have Gaussian turn-on and -off envelopes with a variance of 28.4 ns for all the stretches. The median CNOT error across the device is 7.33e-3 (9.27e-3) for the $c=1$ ($c=1.3$) stretch and the median single qubit gate error is 2.97e-4 (with negligible difference for both stretches) as measured by randomized benchmarking ~\cite{Magesan2011}. 

\begin{figure*}[t!]
	\centering
	\includegraphics[width=\textwidth]{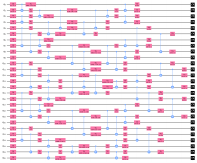}
	\caption{\textbf{Quantum circuit for labeling cosets.} Physical gate representation of the circuit family in Fig.~\ref{Fig:Chip_Circuit}(b) for a set of parameters $\lambda$, $\myVec{\theta}$, and $\myVec{\tilde{\theta}}$.}
	\label{fig:circuit}
\end{figure*}

Fig.~\ref{fig:circuit} shows a physical representation of one of the circuits used to evaluate kernel entries, as shown in Fig.~\ref{Fig:Chip_Circuit}(b) in the main text. The data points are entered via the parameters in the $U_3(\theta, \phi, \lambda) = R_z(\phi) R_y (\theta) R_z (\lambda)$ and $U_2(\phi, \lambda) = U_3(\pi/2, \phi, \lambda)$ unitaries, where $R_y(\alpha) = \exp(-i\alpha Y/2)$ and the equivalent form holds for $R_z$. The total circuit length is 4.6615 $\mu$s, including the 340 ns measurement pulse, for the $c=1$ stretch, and 5.9580 $\mu$s for the $c=1.3$ stretch. Note that the measurement pulse is not scaled in length when implementing zero-noise extrapolation.

\subsection{Training data}
\begin{figure*}[t!]
	\centering
	\includegraphics[width=\textwidth]{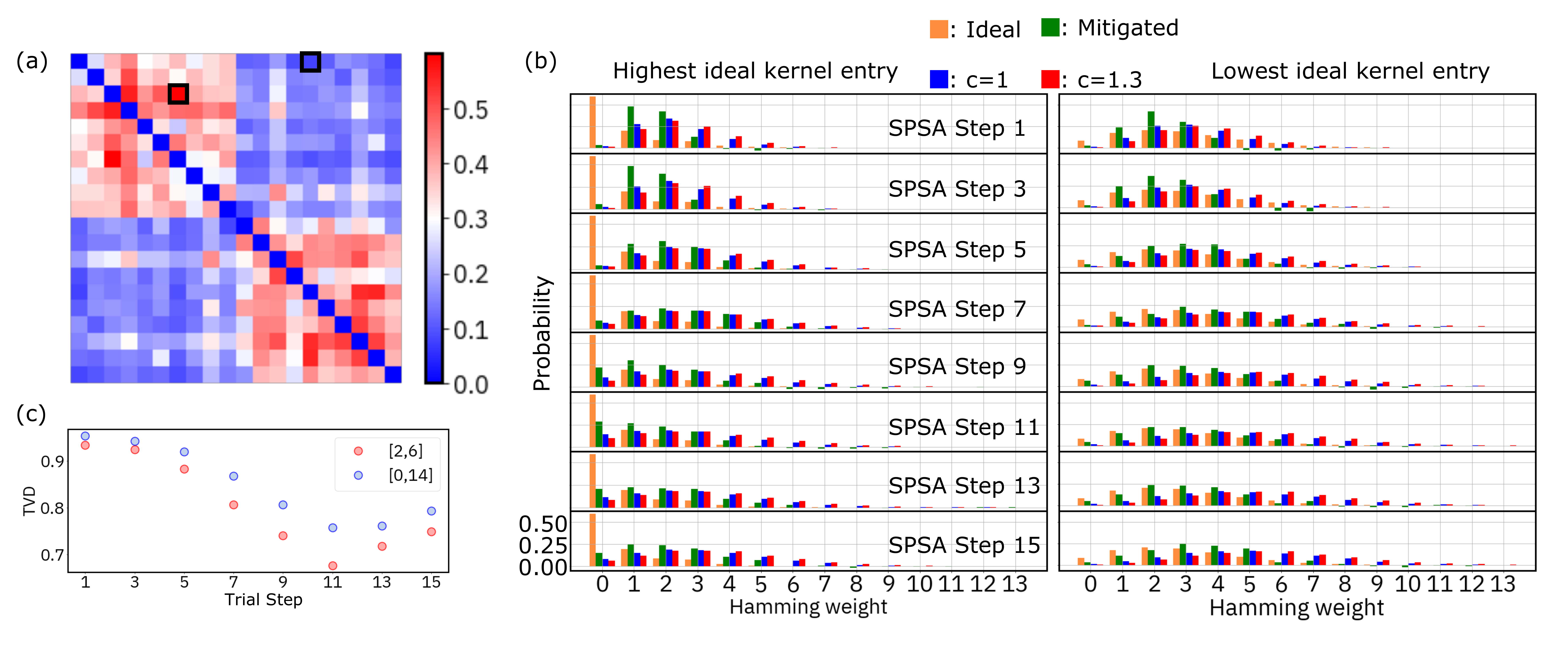}
	\caption{\textbf{Experimental output distributions for training datasets}. (a) Ideal kernel matrix for the training dataset obtained from a noiseless simulation using $\lambda = \pi/2$. The highest (data points 2 and 6) and lowest (data points 0 and 14) entries are highlighted with a black square. (b) Hamming weight distributions of the experimental outputs for the training data points pairs $[2,6]$ (left) and $[0,14]$ (right). The yellow bars show the Hamming weight distribution corresponding to the noiseless simulation of those data pairs using $\lambda = \pi/2$. The stretches $c=1$ (blue bars) and $c=1.3$ (red bars) are combined to extrapolate to the zero-noise limit (green bars). (c) Total variation distance between the $c=1$ experimental output distribution for the two kernel entries highlighted in (a) and the corresponding noiseless distribution as a function of SPSA optimization step.}
	\label{fig:hammings}
\end{figure*}

We offer here more insight into the details of the kernel alignment process with the training dataset. Fig.~\ref{FigExp} in the main text shows experimental kernels at odd steps between 1 and 15 during the kernel alignment process. Those kernels capture the overlap between each pair of training data points $(\myVec{\theta}, \myVec{\tilde{\theta}})$ by measuring the observable  $K_{\myVec{\lambda}}(\myVec{\theta},\myVec{\tilde{\theta}}) = |\bra{0^n }V^\dagger_{\myVec{\lambda}}D^\dagger_{\myVec{\theta}} D_{\tilde{\myVec{\theta}}} V_{\myVec{\lambda}} \ket{0^n}|^2$. Here we look at the output distributions of the quantum computer beyond simply that matrix element for two pairs of data points: the pair $[2, 6]$ and the pair $[0, 14]$. The kernel entries for these two pairs in the ideal case (noiseless simulation for $\lambda = \pi/2$) are marked by black squares in Fig.~\ref{fig:hammings}(a). In Fig.~\ref{fig:hammings}(b) we show the Hamming weight of the experimental output distributions for both pairs of data points using the mitigated $\lambda$ parameter at each of the SPSA steps. The bars show both gate stretch factors ($c=1$ and $c=1.3$) as well as the error mitigated and ideal (noiseless) expectation values for each Hamming weight. We observe that for both kernel entries, the experimental distributions tend to peak at around Hamming weight 2 and 3. For the training pair $[2, 6]$ (Fig.~\ref{fig:hammings}(b) left), however, an increased probability of observing zero Hamming weight progressively develops as the SPSA optimization progresses, peaking at step 11. The pair $[0, 14]$ also shows an increased proximity to the noiseless expectation distribution for the higher SPSA steps measured. Another quantitative approach of the (unmitigated) experimental distribution to the ideal for both pairs of data points can be seen in Fig.~\ref{fig:hammings}(c), where the total variation distance $\mathrm{TVD} = 1/2\sum_i |P_i - Q_i|$ is plotted for both kernel entries as a function of SPSA step.

\begin{figure*}[t!]
	\centering
	\includegraphics[width=\textwidth]{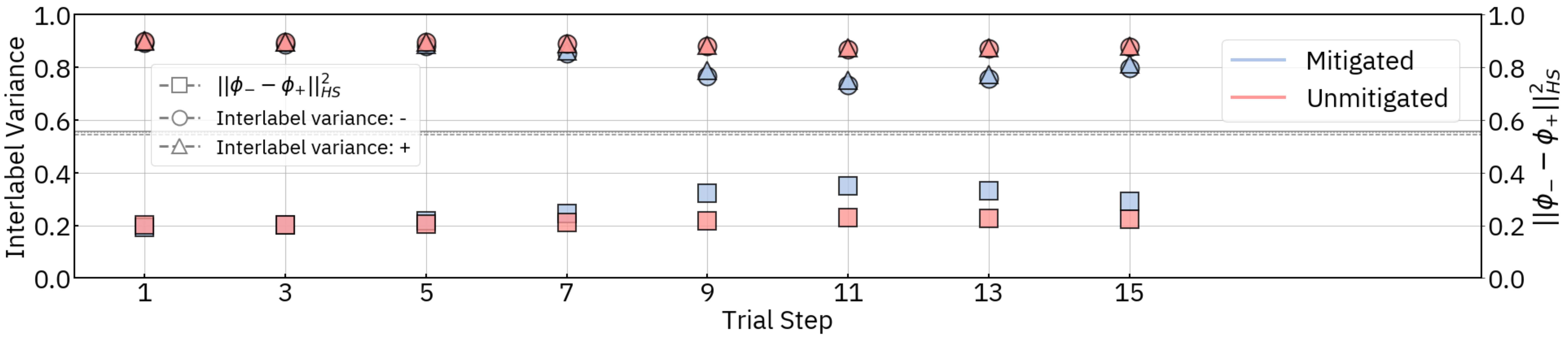}
	\caption{\textbf{Evolution of the mapped training dataset as a function of SPSA trial step}. We look at two metrics related to the data in the feature space as the parameter $\lambda$ evolves with the SPSA optimization. We see that as the optimization progresses, the Hilbert-Schmidt norm of the difference between the centers of mass for each label increases, showing how the datasets get mapped further from each other (square symbols). We also show the interlabel variance, which shows the overall spread of the datasets for each label in the feature space. The variance for both labels decreases as the optimization progresses. For both metrics the mitigated data does better into the optimization than the unmitigated data. The ideal values for each metric, shown by the gray lines, are limited by the level of noise introduced in our datasets ($\epsilon = 0.01$). }
	\label{fig:angles}
\end{figure*}

We can also obtain another view of the convergence of the SPSA optimization by looking at different geometric inspired metrics. By looking at the separation and spread of the training data points in the feature space as the optimization evolves, we can see how the mapping approaches optimal values and how the error mitigation helps with the optimization of the $\lambda$ parameter. In Fig.~\ref{fig:angles} we show two such metrics: the Hilbert-Schmidt norm of the distance between the center of mass for each label subset, and the variance of the data points within each label. 

For the first metric, we define the center of mass for the positive (negative) label data as $\Phi_+ = \frac{\sum_{i \in T_+} \phi(x_i)}{M}$ ($\Phi_- = \frac{\sum_{i \in T_-} \phi(x_i)}{M}$), and define $HS = || \Phi_+ - \Phi_-||_{\mathrm{HS}}^2$, where $M$ is the number of data points for each label. The interlabel variance is defined as $\sigma_+ = \frac{\sum_{i \in T_+} ||\phi(x_i) - \Phi_+||_{\mathrm{HS}}^2}{M}=\frac{\sum_{i \in T_+}\sum_{j \in T_+}  ||\phi(x_i) - \phi(x_j)||_{\mathrm{HS}}^2}{2M^2}$, with an equivalent definition for the negative label set.

Fig.~\ref{fig:angles} shows these two metrics, obtained from the experimentally measured kernels, as a function of the trial step. We see that the interlabel variance decreases with increasing trial step, whereas the Hilbert-Schmidt norm of the difference between the center of masses for each label increases with increasing trial step. The ideal values for each metric are shown as gray lines (solid for the center of mass difference, dashed for the interlabel plus set, dotted for the interlabel minus set). The mitigated approach (blue) shows in all cases a measurable advantage versus the non-mitigated data.

\subsection{Testing data}
\begin{figure*}[t!]
	\centering
	\includegraphics[width=\textwidth]{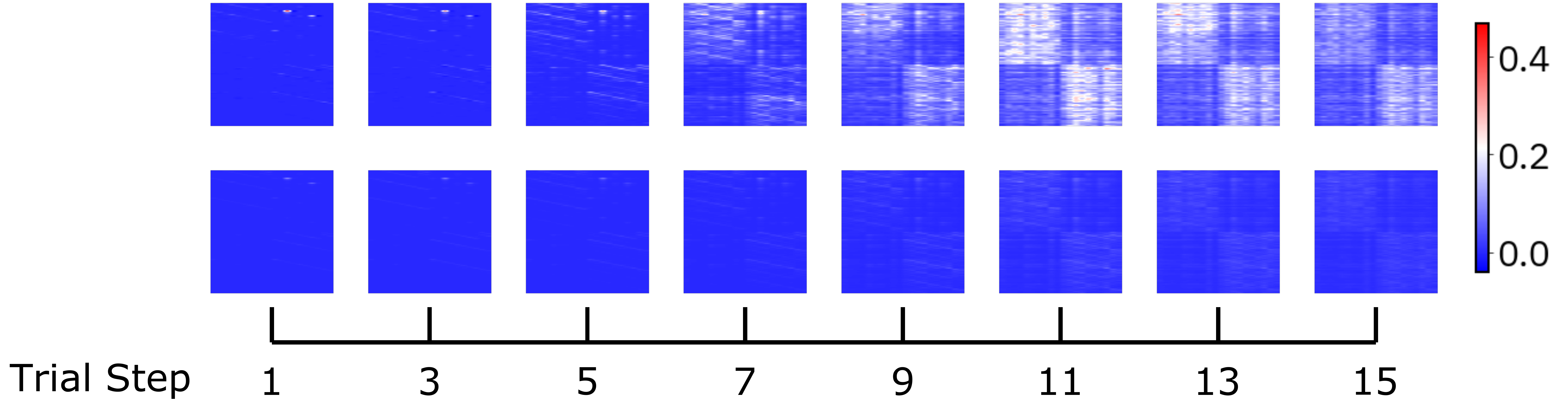}
	\caption{\textbf{Training sets kernels.} Kernel matrices with (upper row) and without (lower row) error mitigation for the test data sets as a function of SPSA step.}
	\label{fig:test_kernels}
\end{figure*}

Similar to the display of the training kernels in the main text, we show in Fig.~\ref{fig:test_kernels} the Gram matrices for the test data for odd trial steps between 1 and 15. The error mitigated matrices (which also use the error mitigated $\lambda$ parameter) show a clearer contrast between labels compared to the unmitigated matrices. In this figure, the training data points (10 per label) correspond to the matrices columns and the test data points (50 per label) correspond to the matrices rows.

\begin{figure*}[t!]
	\centering
	\includegraphics[width=\textwidth]{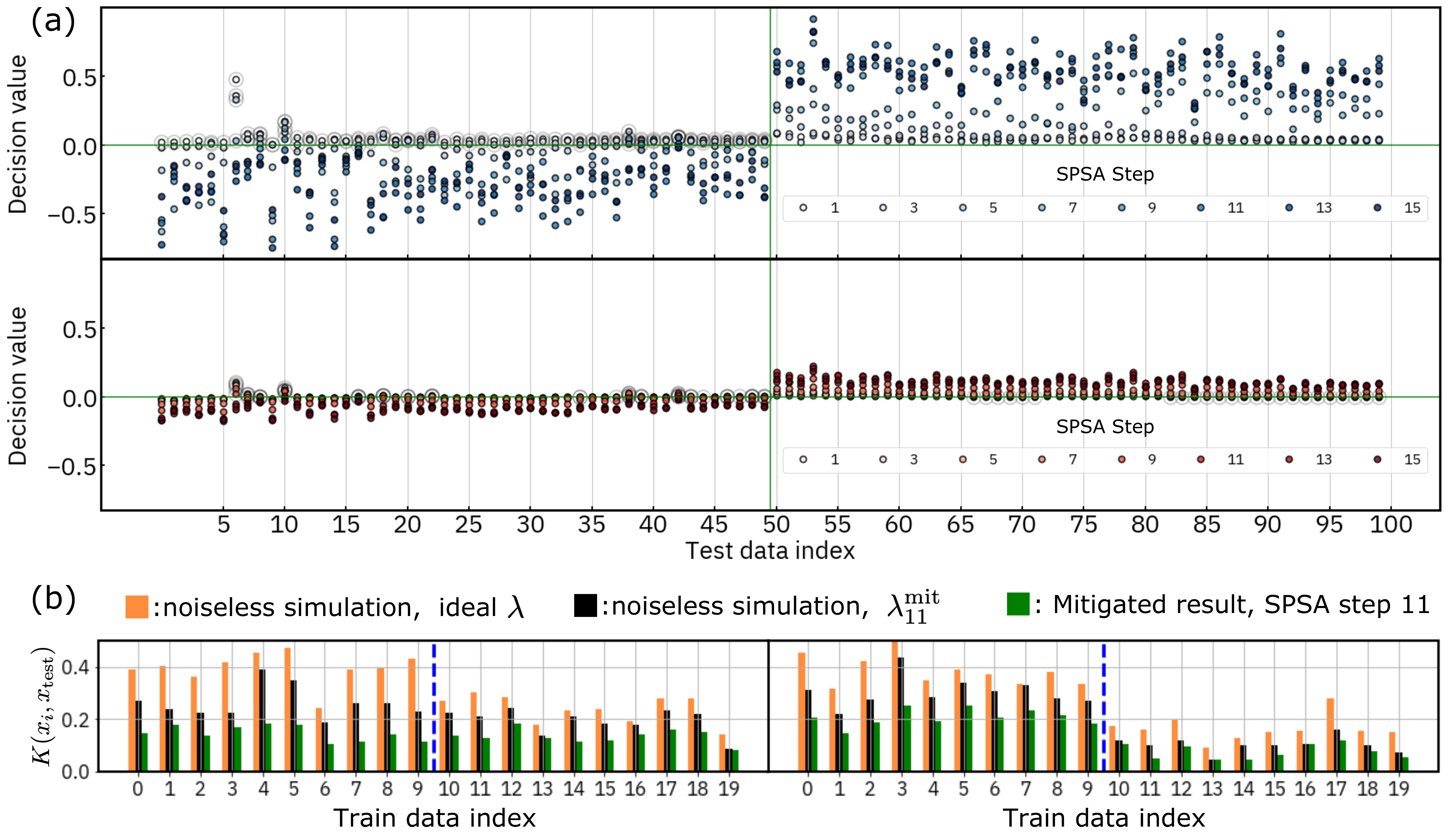}
	\caption{\textbf{A closer look at the classification of the test dataset}.(a) Decision values for each test point classification at each SPSA step for the mitigated (upper panel) and unmitigated (lower panel) approaches. Test data indexes below 50 correspond to negative labels. The increasing margin for the decision values as the SPSA optimization progresses is quite evident for both the mitigated an unmitigated approaches, the former attaining much larger margins as expected. Incorrectly classified data points are highlighted with an outer circle. (b) The overlap of two test data points with negative label (index 42, left; index 5, right) with each of the 20 training data points. Training data point indexes $[0, 9]$ ($[10, 19]$) correspond to negative (positive) label. The dashed blue lines separate the training data points classes. The kernel entries for each training data point are shown for the noiseless simulation for the ideal fiducial state (orange), the noiseless simulation for the experimentally obtained mitigated $\lambda$ at SPSA step 11 (black), and the error mitigated experimental outcomes at SPSA step 11 (green).}
	\label{fig:decision}
\end{figure*}

Beyond simply reporting the classification result of each data point in our test dataset, we can look at the precise value of the decision function in each case, $d(x) = \sum_{i\in SV} y_i \alpha_i K(x_i, x) +b$, where $SV$ is the set of support vectors. This adds valuable information to the classification results and offers interesting insights into the performance of the classifier for the different values of $\lambda$ and for each particular test data point. We show the decision value for each test data point in Fig.~\ref{fig:decision}(a). The top panel (blue symbols) corresponds to the error mitigated data and the bottom panel (red symbols) corresponds to the unmitigated data. Darker symbol colors in each case correspond to later SPSA trial steps. Misclassified data points are circled for more clarity. We see that the mitigated data show consistently larger decision values at the later stages of the SPSA optimization compared to the unmitigated data. There are some points, however, that remain very close to the decision boundary even deep into the SPSA process, as is the case for data point 42. In Fig.~\ref{fig:decision}(b) we show the overlap between two particular test data points, with indexes 42 and 5, and each of the training data points for the SPSA trial step 11. These are two examples of test data points with reasonably low and high values, respectively, for their decision function. Looking at the mitigated results (green bars) in Fig.~\ref{fig:decision}(b), which represent the corresponding row for each test data point in the mitigated Gram matrix for trial step 11 as shown in Fig.~\ref{fig:test_kernels}, we see that the overlap with the training set for point 42 is quite uniform across the training set, independent of the training label, whereas the overlap for point 5 is remarkably larger with the training points with negative label, which results in an easier classification. Note that, of all the training data points, only index 11 is not a support vector for the kernel corresponding to the mitigated SPSA step 11. Noiseless simulations using the ideal value $\lambda=\pi/2$ (orange bars in Fig.~\ref{fig:decision}(b)) show that the classification of test data point 42 is indeed challenging even in the absence of experimental noise, due to its overlap with support vectors of the opposite label, whereas the same computation for test data point 5 shows this latter point is much easier to classify.

\pagebreak

\begin{thebibliography}{66}%
\makeatletter
\providecommand \@ifxundefined [1]{%
 \@ifx{#1\undefined}
}%
\providecommand \@ifnum [1]{%
 \ifnum #1\expandafter \@firstoftwo
 \else \expandafter \@secondoftwo
 \fi
}%
\providecommand \@ifx [1]{%
 \ifx #1\expandafter \@firstoftwo
 \else \expandafter \@secondoftwo
 \fi
}%
\providecommand \natexlab [1]{#1}%
\providecommand \enquote  [1]{``#1''}%
\providecommand \bibnamefont  [1]{#1}%
\providecommand \bibfnamefont [1]{#1}%
\providecommand \citenamefont [1]{#1}%
\providecommand \href@noop [0]{\@secondoftwo}%
\providecommand \href [0]{\begingroup \@sanitize@url \@href}%
\providecommand \@href[1]{\@@startlink{#1}\@@href}%
\providecommand \@@href[1]{\endgroup#1\@@endlink}%
\providecommand \@sanitize@url [0]{\catcode `\\12\catcode `\$12\catcode
  `\&12\catcode `\#12\catcode `\^12\catcode `\_12\catcode `\%12\relax}%
\providecommand \@@startlink[1]{}%
\providecommand \@@endlink[0]{}%
\providecommand \url  [0]{\begingroup\@sanitize@url \@url }%
\providecommand \@url [1]{\endgroup\@href {#1}{\urlprefix }}%
\providecommand \urlprefix  [0]{URL }%
\providecommand \Eprint [0]{\href }%
\providecommand \doibase [0]{http://dx.doi.org/}%
\providecommand \selectlanguage [0]{\@gobble}%
\providecommand \bibinfo  [0]{\@secondoftwo}%
\providecommand \bibfield  [0]{\@secondoftwo}%
\providecommand \translation [1]{[#1]}%
\providecommand \BibitemOpen [0]{}%
\providecommand \bibitemStop [0]{}%
\providecommand \bibitemNoStop [0]{.\EOS\space}%
\providecommand \EOS [0]{\spacefactor3000\relax}%
\providecommand \BibitemShut  [1]{\csname bibitem#1\endcsname}%
\let\auto@bib@innerbib\@empty
\bibitem [{\citenamefont {Boser}\ \emph {et~al.}(1992)\citenamefont {Boser},
  \citenamefont {Guyon},\ and\ \citenamefont {Vapnik}}]{boser1992training}%
  \BibitemOpen
  \bibfield  {author} {\bibinfo {author} {\bibfnamefont {B.~E.}\ \bibnamefont
  {Boser}}, \bibinfo {author} {\bibfnamefont {I.~M.}\ \bibnamefont {Guyon}}, \
  and\ \bibinfo {author} {\bibfnamefont {V.~N.}\ \bibnamefont {Vapnik}},\ }in\
  \href {\doibase 10.1145/130385.130401} {\emph {\bibinfo {booktitle}
  {Proceedings of the Fifth Annual Workshop on Computational Learning
  Theory}}},\ \bibinfo {series and number} {COLT '92}\ (\bibinfo  {publisher}
  {Association for Computing Machinery},\ \bibinfo {address} {New York, NY,
  USA},\ \bibinfo {year} {1992})\ pp.\ \bibinfo {pages} {144--152}\BibitemShut
  {NoStop}%
\bibitem [{\citenamefont {Vapnik}\ \emph {et~al.}(1996)\citenamefont {Vapnik},
  \citenamefont {Golowich},\ and\ \citenamefont
  {Smola}}]{vapnik1996regression}%
  \BibitemOpen
  \bibfield  {author} {\bibinfo {author} {\bibfnamefont {V.}~\bibnamefont
  {Vapnik}}, \bibinfo {author} {\bibfnamefont {S.~E.}\ \bibnamefont
  {Golowich}}, \ and\ \bibinfo {author} {\bibfnamefont {A.}~\bibnamefont
  {Smola}},\ }in\ \href {https://dl.acm.org/doi/abs/10.5555/2998981.2999021}
  {\emph {\bibinfo {booktitle} {Proceedings of the 9th International Conference
  on Neural Information Processing Systems}}},\ \bibinfo {series and number}
  {NIPS'96}\ (\bibinfo  {publisher} {MIT Press},\ \bibinfo {address}
  {Cambridge, MA, USA},\ \bibinfo {year} {1996})\ pp.\ \bibinfo {pages}
  {281--287}\BibitemShut {NoStop}%
\bibitem [{\citenamefont {Smola}\ and\ \citenamefont
  {Sch{\"o}lkopf}(2004)}]{smola2004tutorial}%
  \BibitemOpen
  \bibfield  {author} {\bibinfo {author} {\bibfnamefont {A.~J.}\ \bibnamefont
  {Smola}}\ and\ \bibinfo {author} {\bibfnamefont {B.}~\bibnamefont
  {Sch{\"o}lkopf}},\ }\href
  {https://doi.org/10.1023/B:STCO.0000035301.49549.88} {\bibfield  {journal}
  {\bibinfo  {journal} {Statistics and computing}\ }\textbf {\bibinfo {volume}
  {14}},\ \bibinfo {pages} {199} (\bibinfo {year} {2004})}\BibitemShut
  {NoStop}%
\bibitem [{\citenamefont {Ben-Hur}\ \emph {et~al.}(2002)\citenamefont
  {Ben-Hur}, \citenamefont {Horn}, \citenamefont {Siegelmann},\ and\
  \citenamefont {Vapnik}}]{benhur2002support}%
  \BibitemOpen
  \bibfield  {author} {\bibinfo {author} {\bibfnamefont {A.}~\bibnamefont
  {Ben-Hur}}, \bibinfo {author} {\bibfnamefont {D.}~\bibnamefont {Horn}},
  \bibinfo {author} {\bibfnamefont {H.~T.}\ \bibnamefont {Siegelmann}}, \ and\
  \bibinfo {author} {\bibfnamefont {V.}~\bibnamefont {Vapnik}},\ }\href
  {https://dl.acm.org/doi/abs/10.5555/944790.944807} {\bibfield  {journal}
  {\bibinfo  {journal} {J. Mach. Learn. Res.}\ }\textbf {\bibinfo {volume}
  {2}},\ \bibinfo {pages} {125} (\bibinfo {year} {2002})}\BibitemShut {NoStop}%
\bibitem [{\citenamefont {Girolami}(2002)}]{girolami2002mercer}%
  \BibitemOpen
  \bibfield  {author} {\bibinfo {author} {\bibfnamefont {M.}~\bibnamefont
  {Girolami}},\ }\href {\doibase 10.1109/TNN.2002.1000150} {\bibfield
  {journal} {\bibinfo  {journal} {IEEE Transactions on Neural Networks}\
  }\textbf {\bibinfo {volume} {13}},\ \bibinfo {pages} {780} (\bibinfo {year}
  {2002})}\BibitemShut {NoStop}%
\bibitem [{\citenamefont {Lai}\ and\ \citenamefont
  {Fyfe}(2000)}]{lai2000kernel}%
  \BibitemOpen
  \bibfield  {author} {\bibinfo {author} {\bibfnamefont {P.~L.}\ \bibnamefont
  {Lai}}\ and\ \bibinfo {author} {\bibfnamefont {C.}~\bibnamefont {Fyfe}},\
  }\href {https://doi.org/10.1142/S012906570000034X} {\bibfield  {journal}
  {\bibinfo  {journal} {International Journal of Neural Systems}\ }\textbf
  {\bibinfo {volume} {10}},\ \bibinfo {pages} {365} (\bibinfo {year}
  {2000})}\BibitemShut {NoStop}%
\bibitem [{\citenamefont {Liu}\ \emph {et~al.}(2011)\citenamefont {Liu},
  \citenamefont {Principe},\ and\ \citenamefont {Haykin}}]{liu2011kernel}%
  \BibitemOpen
  \bibfield  {author} {\bibinfo {author} {\bibfnamefont {W.}~\bibnamefont
  {Liu}}, \bibinfo {author} {\bibfnamefont {J.~C.}\ \bibnamefont {Principe}}, \
  and\ \bibinfo {author} {\bibfnamefont {S.}~\bibnamefont {Haykin}},\ }\href
  {https://doi.org/10.1002/9780470608593} {\emph {\bibinfo {title} {Kernel
  adaptive filtering: a comprehensive introduction}}},\ Vol.~\bibinfo {volume}
  {57}\ (\bibinfo  {publisher} {John Wiley \& Sons},\ \bibinfo {year}
  {2011})\BibitemShut {NoStop}%
\bibitem [{\citenamefont {Vapnik}(2013)}]{vapnik2013nature}%
  \BibitemOpen
  \bibfield  {author} {\bibinfo {author} {\bibfnamefont {V.}~\bibnamefont
  {Vapnik}},\ }\href {https://books.google.com/books?id=EqgACAAAQBAJ} {\emph
  {\bibinfo {title} {The Nature of Statistical Learning Theory}}},\ Information
  Science and Statistics\ (\bibinfo  {publisher} {Springer New York},\ \bibinfo
  {year} {2013})\BibitemShut {NoStop}%
\bibitem [{\citenamefont {Havl{\'\i}{\v{c}}ek}\ \emph
  {et~al.}(2019)\citenamefont {Havl{\'\i}{\v{c}}ek}, \citenamefont
  {C{\'o}rcoles}, \citenamefont {Temme}, \citenamefont {Harrow}, \citenamefont
  {Kandala}, \citenamefont {Chow},\ and\ \citenamefont
  {Gambetta}}]{havlivcek2019supervised}%
  \BibitemOpen
  \bibfield  {author} {\bibinfo {author} {\bibfnamefont {V.}~\bibnamefont
  {Havl{\'\i}{\v{c}}ek}}, \bibinfo {author} {\bibfnamefont {A.~D.}\
  \bibnamefont {C{\'o}rcoles}}, \bibinfo {author} {\bibfnamefont
  {K.}~\bibnamefont {Temme}}, \bibinfo {author} {\bibfnamefont {A.~W.}\
  \bibnamefont {Harrow}}, \bibinfo {author} {\bibfnamefont {A.}~\bibnamefont
  {Kandala}}, \bibinfo {author} {\bibfnamefont {J.~M.}\ \bibnamefont {Chow}}, \
  and\ \bibinfo {author} {\bibfnamefont {J.~M.}\ \bibnamefont {Gambetta}},\
  }\href {https://doi.org/10.1038/s41586-019-0980-2} {\bibfield  {journal}
  {\bibinfo  {journal} {Nature}\ }\textbf {\bibinfo {volume} {567}},\ \bibinfo
  {pages} {209} (\bibinfo {year} {2019})}\BibitemShut {NoStop}%
\bibitem [{\citenamefont {Schuld}\ and\ \citenamefont
  {Killoran}(2019)}]{schuld2019quantum}%
  \BibitemOpen
  \bibfield  {author} {\bibinfo {author} {\bibfnamefont {M.}~\bibnamefont
  {Schuld}}\ and\ \bibinfo {author} {\bibfnamefont {N.}~\bibnamefont
  {Killoran}},\ }\href {\doibase 10.1103/PhysRevLett.122.040504} {\bibfield
  {journal} {\bibinfo  {journal} {Phys. Rev. Lett.}\ }\textbf {\bibinfo
  {volume} {122}},\ \bibinfo {pages} {040504} (\bibinfo {year}
  {2019})}\BibitemShut {NoStop}%
\bibitem [{\citenamefont {Mitarai}\ \emph {et~al.}(2018)\citenamefont
  {Mitarai}, \citenamefont {Negoro}, \citenamefont {Kitagawa},\ and\
  \citenamefont {Fujii}}]{mitarai2018quantum}%
  \BibitemOpen
  \bibfield  {author} {\bibinfo {author} {\bibfnamefont {K.}~\bibnamefont
  {Mitarai}}, \bibinfo {author} {\bibfnamefont {M.}~\bibnamefont {Negoro}},
  \bibinfo {author} {\bibfnamefont {M.}~\bibnamefont {Kitagawa}}, \ and\
  \bibinfo {author} {\bibfnamefont {K.}~\bibnamefont {Fujii}},\ }\href
  {\doibase 10.1103/PhysRevA.98.032309} {\bibfield  {journal} {\bibinfo
  {journal} {Phys. Rev. A}\ }\textbf {\bibinfo {volume} {98}},\ \bibinfo
  {pages} {032309} (\bibinfo {year} {2018})}\BibitemShut {NoStop}%
\bibitem [{\citenamefont {Farhi}\ and\ \citenamefont
  {Neven}(2018)}]{farhi2018classification}%
  \BibitemOpen
  \bibfield  {author} {\bibinfo {author} {\bibfnamefont {E.}~\bibnamefont
  {Farhi}}\ and\ \bibinfo {author} {\bibfnamefont {H.}~\bibnamefont {Neven}},\
  }\href {https://arxiv.org/abs/1802.06002} {\bibfield  {journal} {\bibinfo
  {journal} {arXiv:1802.06002}\ } (\bibinfo {year} {2018})}\BibitemShut
  {NoStop}%
\bibitem [{\citenamefont {Grant}\ \emph {et~al.}(2018)\citenamefont {Grant},
  \citenamefont {Benedetti}, \citenamefont {Cao}, \citenamefont {Hallam},
  \citenamefont {Lockhart}, \citenamefont {Stojevic}, \citenamefont {Green},\
  and\ \citenamefont {Severini}}]{grant2018hierarchical}%
  \BibitemOpen
  \bibfield  {author} {\bibinfo {author} {\bibfnamefont {E.}~\bibnamefont
  {Grant}}, \bibinfo {author} {\bibfnamefont {M.}~\bibnamefont {Benedetti}},
  \bibinfo {author} {\bibfnamefont {S.}~\bibnamefont {Cao}}, \bibinfo {author}
  {\bibfnamefont {A.}~\bibnamefont {Hallam}}, \bibinfo {author} {\bibfnamefont
  {J.}~\bibnamefont {Lockhart}}, \bibinfo {author} {\bibfnamefont
  {V.}~\bibnamefont {Stojevic}}, \bibinfo {author} {\bibfnamefont {A.~G.}\
  \bibnamefont {Green}}, \ and\ \bibinfo {author} {\bibfnamefont
  {S.}~\bibnamefont {Severini}},\ }\href
  {https://doi.org/10.1038/s41534-018-0116-9} {\bibfield  {journal} {\bibinfo
  {journal} {npj Quantum Information}\ }\textbf {\bibinfo {volume} {4}},\
  \bibinfo {pages} {1} (\bibinfo {year} {2018})}\BibitemShut {NoStop}%
\bibitem [{\citenamefont {Schuld}\ \emph {et~al.}(2020)\citenamefont {Schuld},
  \citenamefont {Bocharov}, \citenamefont {Svore},\ and\ \citenamefont
  {Wiebe}}]{schuld2020circuit}%
  \BibitemOpen
  \bibfield  {author} {\bibinfo {author} {\bibfnamefont {M.}~\bibnamefont
  {Schuld}}, \bibinfo {author} {\bibfnamefont {A.}~\bibnamefont {Bocharov}},
  \bibinfo {author} {\bibfnamefont {K.~M.}\ \bibnamefont {Svore}}, \ and\
  \bibinfo {author} {\bibfnamefont {N.}~\bibnamefont {Wiebe}},\ }\href
  {\doibase 10.1103/PhysRevA.101.032308} {\bibfield  {journal} {\bibinfo
  {journal} {Phys. Rev. A}\ }\textbf {\bibinfo {volume} {101}},\ \bibinfo
  {pages} {032308} (\bibinfo {year} {2020})}\BibitemShut {NoStop}%
\bibitem [{\citenamefont {Benedetti}\ \emph {et~al.}(2019)\citenamefont
  {Benedetti}, \citenamefont {Lloyd}, \citenamefont {Sack},\ and\ \citenamefont
  {Fiorentini}}]{benedetti2019parameterized}%
  \BibitemOpen
  \bibfield  {author} {\bibinfo {author} {\bibfnamefont {M.}~\bibnamefont
  {Benedetti}}, \bibinfo {author} {\bibfnamefont {E.}~\bibnamefont {Lloyd}},
  \bibinfo {author} {\bibfnamefont {S.}~\bibnamefont {Sack}}, \ and\ \bibinfo
  {author} {\bibfnamefont {M.}~\bibnamefont {Fiorentini}},\ }\href
  {https://doi.org/10.1088/2058-9565/ab4eb5} {\bibfield  {journal} {\bibinfo
  {journal} {Quantum Science and Technology}\ }\textbf {\bibinfo {volume}
  {4}},\ \bibinfo {pages} {043001} (\bibinfo {year} {2019})}\BibitemShut
  {NoStop}%
\bibitem [{\citenamefont {Liu}\ \emph {et~al.}(2021)\citenamefont {Liu},
  \citenamefont {Arunachalam},\ and\ \citenamefont {Temme}}]{liu2020rigorous}%
  \BibitemOpen
  \bibfield  {author} {\bibinfo {author} {\bibfnamefont {Y.}~\bibnamefont
  {Liu}}, \bibinfo {author} {\bibfnamefont {S.}~\bibnamefont {Arunachalam}}, \
  and\ \bibinfo {author} {\bibfnamefont {K.}~\bibnamefont {Temme}},\ }\href
  {https://doi.org/10.1038/s41567-021-01287-z} {\bibfield  {journal} {\bibinfo
  {journal} {Nature Physics}\ }\textbf {\bibinfo {volume} {17}},\ \bibinfo
  {pages} {1013} (\bibinfo {year} {2021})}\BibitemShut {NoStop}%
\bibitem [{\citenamefont {Diaconis}(1988)}]{diaconis1988group}%
  \BibitemOpen
  \bibfield  {author} {\bibinfo {author} {\bibfnamefont {P.}~\bibnamefont
  {Diaconis}},\ }\href@noop {} {\bibfield  {journal} {\bibinfo  {journal}
  {Lecture notes-monograph series}\ }\textbf {\bibinfo {volume} {11}},\
  \bibinfo {pages} {i} (\bibinfo {year} {1988})}\BibitemShut {NoStop}%
\bibitem [{\citenamefont {Kondor}\ and\ \citenamefont
  {Barbosa}(2010)}]{kondor2010ranking}%
  \BibitemOpen
  \bibfield  {author} {\bibinfo {author} {\bibfnamefont {R.}~\bibnamefont
  {Kondor}}\ and\ \bibinfo {author} {\bibfnamefont {M.~S.}\ \bibnamefont
  {Barbosa}},\ }in\ \href
  {http://galton.uchicago.edu/~lekheng/meetings/mathofranking/ref/kondor.pdf}
  {\emph {\bibinfo {booktitle} {COLT}}}\ (\bibinfo {year} {2010})\ pp.\
  \bibinfo {pages} {451--463}\BibitemShut {NoStop}%
\bibitem [{\citenamefont {Kondor}(2008)}]{kondor2008group}%
  \BibitemOpen
  \bibfield  {author} {\bibinfo {author} {\bibfnamefont {I.~R.}\ \bibnamefont
  {Kondor}},\ }\emph {\bibinfo {title} {Group theoretical methods in machine
  learning}},\ \href
  {https://search.proquest.com/openview/d8ae7d9e47c87bb0acb05bf06ae8b6aa/1?pq-origsite=gscholar&cbl=18750&diss=y}
  {Ph.D. thesis} (\bibinfo {year} {2008})\BibitemShut {NoStop}%
\bibitem [{\citenamefont {Jiao}\ and\ \citenamefont
  {Vert}(2015)}]{jiao2015kendall}%
  \BibitemOpen
  \bibfield  {author} {\bibinfo {author} {\bibfnamefont {Y.}~\bibnamefont
  {Jiao}}\ and\ \bibinfo {author} {\bibfnamefont {J.-P.}\ \bibnamefont
  {Vert}},\ }in\ \href {https://dl.acm.org/doi/abs/10.5555/3045118.3045324}
  {\emph {\bibinfo {booktitle} {International Conference on Machine
  Learning}}}\ (\bibinfo {organization} {PMLR},\ \bibinfo {year} {2015})\ pp.\
  \bibinfo {pages} {1935--1944}\BibitemShut {NoStop}%
\bibitem [{\citenamefont {Holevo}(1979)}]{holevo1979covariant}%
  \BibitemOpen
  \bibfield  {author} {\bibinfo {author} {\bibfnamefont {A.}~\bibnamefont
  {Holevo}},\ }\href {https://doi.org/10.1016/0034-4877(79)90072-7} {\bibfield
  {journal} {\bibinfo  {journal} {Reports on mathematical physics}\ }\textbf
  {\bibinfo {volume} {16}},\ \bibinfo {pages} {385} (\bibinfo {year}
  {1979})}\BibitemShut {NoStop}%
\bibitem [{\citenamefont {Serre}(1977)}]{serre1977linear}%
  \BibitemOpen
  \bibfield  {author} {\bibinfo {author} {\bibfnamefont {J.-P.}\ \bibnamefont
  {Serre}},\ }\href {https://doi.org/10.1007/978-1-4684-9458-7} {\emph
  {\bibinfo {title} {Linear representations of finite groups}}},\ Vol.~\bibinfo
  {volume} {42}\ (\bibinfo  {publisher} {Springer},\ \bibinfo {year}
  {1977})\BibitemShut {NoStop}%
\bibitem [{\citenamefont {Rosen}(2011)}]{rosen2011elementary}%
  \BibitemOpen
  \bibfield  {author} {\bibinfo {author} {\bibfnamefont {K.~H.}\ \bibnamefont
  {Rosen}},\ }\href {https://books.google.com/books?id=JqycRAAACAAJ} {\emph
  {\bibinfo {title} {Elementary number theory}}}\ (\bibinfo  {publisher}
  {Pearson Education London},\ \bibinfo {year} {2011})\BibitemShut {NoStop}%
\bibitem [{\citenamefont {Lloyd}\ \emph {et~al.}(2020)\citenamefont {Lloyd},
  \citenamefont {Schuld}, \citenamefont {Ijaz}, \citenamefont {Izaac},\ and\
  \citenamefont {Killoran}}]{lloyd2020quantum}%
  \BibitemOpen
  \bibfield  {author} {\bibinfo {author} {\bibfnamefont {S.}~\bibnamefont
  {Lloyd}}, \bibinfo {author} {\bibfnamefont {M.}~\bibnamefont {Schuld}},
  \bibinfo {author} {\bibfnamefont {A.}~\bibnamefont {Ijaz}}, \bibinfo {author}
  {\bibfnamefont {J.}~\bibnamefont {Izaac}}, \ and\ \bibinfo {author}
  {\bibfnamefont {N.}~\bibnamefont {Killoran}},\ }\href
  {https://arxiv.org/abs/2001.03622} {\bibfield  {journal} {\bibinfo  {journal}
  {arXiv:2001.03622}\ } (\bibinfo {year} {2020})}\BibitemShut {NoStop}%
\bibitem [{\citenamefont {Otten}\ \emph {et~al.}(2020)\citenamefont {Otten},
  \citenamefont {Goumiri}, \citenamefont {Priest}, \citenamefont {Chapline},\
  and\ \citenamefont {Schneider}}]{otten2020quantum}%
  \BibitemOpen
  \bibfield  {author} {\bibinfo {author} {\bibfnamefont {M.}~\bibnamefont
  {Otten}}, \bibinfo {author} {\bibfnamefont {I.~R.}\ \bibnamefont {Goumiri}},
  \bibinfo {author} {\bibfnamefont {B.~W.}\ \bibnamefont {Priest}}, \bibinfo
  {author} {\bibfnamefont {G.~F.}\ \bibnamefont {Chapline}}, \ and\ \bibinfo
  {author} {\bibfnamefont {M.~D.}\ \bibnamefont {Schneider}},\ }\href
  {https://arxiv.org/abs/2004.11280} {\bibfield  {journal} {\bibinfo  {journal}
  {arXiv:2004.11280}\ } (\bibinfo {year} {2020})}\BibitemShut {NoStop}%
\bibitem [{\citenamefont {Cristianini}\ \emph {et~al.}(2001)\citenamefont
  {Cristianini}, \citenamefont {Shawe-Taylor}, \citenamefont {Elisseeff},\ and\
  \citenamefont {Kandola}}]{cristianini2001kernel}%
  \BibitemOpen
  \bibfield  {author} {\bibinfo {author} {\bibfnamefont {N.}~\bibnamefont
  {Cristianini}}, \bibinfo {author} {\bibfnamefont {J.}~\bibnamefont
  {Shawe-Taylor}}, \bibinfo {author} {\bibfnamefont {A.}~\bibnamefont
  {Elisseeff}}, \ and\ \bibinfo {author} {\bibfnamefont {J.}~\bibnamefont
  {Kandola}},\ }in\ \href
  {https://proceedings.neurips.cc/paper/2001/file/1f71e393b3809197ed66df836fe833e5-Paper.pdf}
  {\emph {\bibinfo {booktitle} {Advances in Neural Information Processing
  Systems 14}}}\ (\bibinfo {organization} {Citeseer},\ \bibinfo {year}
  {2001})\BibitemShut {NoStop}%
\bibitem [{\citenamefont {Cortes}\ \emph {et~al.}(2012)\citenamefont {Cortes},
  \citenamefont {Mohri},\ and\ \citenamefont
  {Rostamizadeh}}]{cortes2012algorithms}%
  \BibitemOpen
  \bibfield  {author} {\bibinfo {author} {\bibfnamefont {C.}~\bibnamefont
  {Cortes}}, \bibinfo {author} {\bibfnamefont {M.}~\bibnamefont {Mohri}}, \
  and\ \bibinfo {author} {\bibfnamefont {A.}~\bibnamefont {Rostamizadeh}},\
  }\href {https://dl.acm.org/doi/10.5555/2503308.2188413} {\bibfield  {journal}
  {\bibinfo  {journal} {The Journal of Machine Learning Research}\ }\textbf
  {\bibinfo {volume} {13}},\ \bibinfo {pages} {795} (\bibinfo {year}
  {2012})}\BibitemShut {NoStop}%
\bibitem [{\citenamefont {Bullins}\ \emph {et~al.}(2017)\citenamefont
  {Bullins}, \citenamefont {Zhang},\ and\ \citenamefont
  {Zhang}}]{bullins2017not}%
  \BibitemOpen
  \bibfield  {author} {\bibinfo {author} {\bibfnamefont {B.}~\bibnamefont
  {Bullins}}, \bibinfo {author} {\bibfnamefont {C.}~\bibnamefont {Zhang}}, \
  and\ \bibinfo {author} {\bibfnamefont {Y.}~\bibnamefont {Zhang}},\ }\href
  {https://arxiv.org/abs/1710.10230} {\bibfield  {journal} {\bibinfo  {journal}
  {arXiv:1710.10230}\ } (\bibinfo {year} {2017})}\BibitemShut {NoStop}%
\bibitem [{\citenamefont {Chamberland}\ \emph {et~al.}(2020)\citenamefont
  {Chamberland}, \citenamefont {Zhu}, \citenamefont {Yoder}, \citenamefont
  {Hertzberg},\ and\ \citenamefont {Cross}}]{chamberland2020topo}%
  \BibitemOpen
  \bibfield  {author} {\bibinfo {author} {\bibfnamefont {C.}~\bibnamefont
  {Chamberland}}, \bibinfo {author} {\bibfnamefont {G.}~\bibnamefont {Zhu}},
  \bibinfo {author} {\bibfnamefont {T.~J.}\ \bibnamefont {Yoder}}, \bibinfo
  {author} {\bibfnamefont {J.~B.}\ \bibnamefont {Hertzberg}}, \ and\ \bibinfo
  {author} {\bibfnamefont {A.~W.}\ \bibnamefont {Cross}},\ }\href {\doibase
  10.1103/PhysRevX.10.011022} {\bibfield  {journal} {\bibinfo  {journal} {Phys.
  Rev. X}\ }\textbf {\bibinfo {volume} {10}},\ \bibinfo {pages} {011022}
  (\bibinfo {year} {2020})}\BibitemShut {NoStop}%
\bibitem [{\citenamefont {Jurcevic}\ \emph {et~al.}(2021)\citenamefont
  {Jurcevic}, \citenamefont {Javadi-Abhari}, \citenamefont {Bishop},
  \citenamefont {Lauer}, \citenamefont {Bogorin}, \citenamefont {Brink},
  \citenamefont {Capelluto}, \citenamefont {G{\"u}nl{\"u}k}, \citenamefont
  {Itoko}, \citenamefont {Kanazawa} \emph
  {et~al.}}]{jurcevic2021demonstration}%
  \BibitemOpen
  \bibfield  {author} {\bibinfo {author} {\bibfnamefont {P.}~\bibnamefont
  {Jurcevic}}, \bibinfo {author} {\bibfnamefont {A.}~\bibnamefont
  {Javadi-Abhari}}, \bibinfo {author} {\bibfnamefont {L.~S.}\ \bibnamefont
  {Bishop}}, \bibinfo {author} {\bibfnamefont {I.}~\bibnamefont {Lauer}},
  \bibinfo {author} {\bibfnamefont {D.~F.}\ \bibnamefont {Bogorin}}, \bibinfo
  {author} {\bibfnamefont {M.}~\bibnamefont {Brink}}, \bibinfo {author}
  {\bibfnamefont {L.}~\bibnamefont {Capelluto}}, \bibinfo {author}
  {\bibfnamefont {O.}~\bibnamefont {G{\"u}nl{\"u}k}}, \bibinfo {author}
  {\bibfnamefont {T.}~\bibnamefont {Itoko}}, \bibinfo {author} {\bibfnamefont
  {N.}~\bibnamefont {Kanazawa}},  \emph {et~al.},\ }\href
  {https://doi.org/10.1088/2058-9565/abe519} {\bibfield  {journal} {\bibinfo
  {journal} {Quantum Science and Technology}\ }\textbf {\bibinfo {volume}
  {6}},\ \bibinfo {pages} {025020} (\bibinfo {year} {2021})}\BibitemShut
  {NoStop}%
\bibitem [{\citenamefont {Gottesman}(1997)}]{gottesman1997stabilizer}%
  \BibitemOpen
  \bibfield  {author} {\bibinfo {author} {\bibfnamefont {D.}~\bibnamefont
  {Gottesman}},\ }\href {https://arxiv.org/abs/quant-ph/9705052} {\emph
  {\bibinfo {title} {Stabilizer codes and quantum error correction}}}\
  (\bibinfo  {publisher} {California Institute of Technology},\ \bibinfo {year}
  {1997})\BibitemShut {NoStop}%
\bibitem [{\citenamefont {Hein}\ \emph {et~al.}(2006)\citenamefont {Hein},
  \citenamefont {D{\"u}r}, \citenamefont {Eisert}, \citenamefont {Raussendorf},
  \citenamefont {Nest},\ and\ \citenamefont {Briegel}}]{hein2006entanglement}%
  \BibitemOpen
  \bibfield  {author} {\bibinfo {author} {\bibfnamefont {M.}~\bibnamefont
  {Hein}}, \bibinfo {author} {\bibfnamefont {W.}~\bibnamefont {D{\"u}r}},
  \bibinfo {author} {\bibfnamefont {J.}~\bibnamefont {Eisert}}, \bibinfo
  {author} {\bibfnamefont {R.}~\bibnamefont {Raussendorf}}, \bibinfo {author}
  {\bibfnamefont {M.}~\bibnamefont {Nest}}, \ and\ \bibinfo {author}
  {\bibfnamefont {H.-J.}\ \bibnamefont {Briegel}},\ }\href
  {https://arxiv.org/abs/quant-ph/0602096} {\bibfield  {journal} {\bibinfo
  {journal} {quant-ph/0602096}\ } (\bibinfo {year} {2006})}\BibitemShut
  {NoStop}%
\bibitem [{\citenamefont {Temme}\ \emph {et~al.}(2017)\citenamefont {Temme},
  \citenamefont {Bravyi},\ and\ \citenamefont {Gambetta}}]{temme2017error}%
  \BibitemOpen
  \bibfield  {author} {\bibinfo {author} {\bibfnamefont {K.}~\bibnamefont
  {Temme}}, \bibinfo {author} {\bibfnamefont {S.}~\bibnamefont {Bravyi}}, \
  and\ \bibinfo {author} {\bibfnamefont {J.~M.}\ \bibnamefont {Gambetta}},\
  }\href {\doibase 10.1103/PhysRevLett.119.180509} {\bibfield  {journal}
  {\bibinfo  {journal} {Phys. Rev. Lett.}\ }\textbf {\bibinfo {volume} {119}},\
  \bibinfo {pages} {180509} (\bibinfo {year} {2017})}\BibitemShut {NoStop}%
\bibitem [{\citenamefont {Li}\ and\ \citenamefont
  {Benjamin}(2017)}]{li2017efficient}%
  \BibitemOpen
  \bibfield  {author} {\bibinfo {author} {\bibfnamefont {Y.}~\bibnamefont
  {Li}}\ and\ \bibinfo {author} {\bibfnamefont {S.~C.}\ \bibnamefont
  {Benjamin}},\ }\href {\doibase 10.1103/PhysRevX.7.021050} {\bibfield
  {journal} {\bibinfo  {journal} {Phys. Rev. X}\ }\textbf {\bibinfo {volume}
  {7}},\ \bibinfo {pages} {021050} (\bibinfo {year} {2017})}\BibitemShut
  {NoStop}%
\bibitem [{\citenamefont {Kandala}\ \emph {et~al.}(2019)\citenamefont
  {Kandala}, \citenamefont {Temme}, \citenamefont {C{\'o}rcoles}, \citenamefont
  {Mezzacapo}, \citenamefont {Chow},\ and\ \citenamefont
  {Gambetta}}]{kandala2019error}%
  \BibitemOpen
  \bibfield  {author} {\bibinfo {author} {\bibfnamefont {A.}~\bibnamefont
  {Kandala}}, \bibinfo {author} {\bibfnamefont {K.}~\bibnamefont {Temme}},
  \bibinfo {author} {\bibfnamefont {A.~D.}\ \bibnamefont {C{\'o}rcoles}},
  \bibinfo {author} {\bibfnamefont {A.}~\bibnamefont {Mezzacapo}}, \bibinfo
  {author} {\bibfnamefont {J.~M.}\ \bibnamefont {Chow}}, \ and\ \bibinfo
  {author} {\bibfnamefont {J.~M.}\ \bibnamefont {Gambetta}},\ }\href
  {https://doi.org/10.1038/s41586-019-1040-7} {\bibfield  {journal} {\bibinfo
  {journal} {Nature}\ }\textbf {\bibinfo {volume} {567}},\ \bibinfo {pages}
  {491} (\bibinfo {year} {2019})}\BibitemShut {NoStop}%
\bibitem [{\citenamefont {Huang}\ \emph {et~al.}(2009)\citenamefont {Huang},
  \citenamefont {Guestrin},\ and\ \citenamefont {Guibas}}]{huang2009fourier}%
  \BibitemOpen
  \bibfield  {author} {\bibinfo {author} {\bibfnamefont {J.}~\bibnamefont
  {Huang}}, \bibinfo {author} {\bibfnamefont {C.}~\bibnamefont {Guestrin}}, \
  and\ \bibinfo {author} {\bibfnamefont {L.}~\bibnamefont {Guibas}},\
  }\href@noop {} {\bibfield  {journal} {\bibinfo  {journal} {Journal of Machine
  Learning Research}\ }\textbf {\bibinfo {volume} {10}} (\bibinfo {year}
  {2009})}\BibitemShut {NoStop}%
\bibitem [{\citenamefont {Bravyi}\ \emph
  {et~al.}(2021{\natexlab{a}})\citenamefont {Bravyi}, \citenamefont {Gosset},\
  and\ \citenamefont {Movassagh}}]{bravyi2021classical}%
  \BibitemOpen
  \bibfield  {author} {\bibinfo {author} {\bibfnamefont {S.}~\bibnamefont
  {Bravyi}}, \bibinfo {author} {\bibfnamefont {D.}~\bibnamefont {Gosset}}, \
  and\ \bibinfo {author} {\bibfnamefont {R.}~\bibnamefont {Movassagh}},\ }\href
  {https://doi.org/10.1038/s41567-020-01109-8} {\bibfield  {journal} {\bibinfo
  {journal} {Nature Physics}\ }\textbf {\bibinfo {volume} {17}},\ \bibinfo
  {pages} {337} (\bibinfo {year} {2021}{\natexlab{a}})}\BibitemShut {NoStop}%
\bibitem [{\citenamefont {Bravyi}\ \emph
  {et~al.}(2021{\natexlab{b}})\citenamefont {Bravyi}, \citenamefont {Sheldon},
  \citenamefont {Kandala}, \citenamefont {Mckay},\ and\ \citenamefont
  {Gambetta}}]{bravyi2021mitigating}%
  \BibitemOpen
  \bibfield  {author} {\bibinfo {author} {\bibfnamefont {S.}~\bibnamefont
  {Bravyi}}, \bibinfo {author} {\bibfnamefont {S.}~\bibnamefont {Sheldon}},
  \bibinfo {author} {\bibfnamefont {A.}~\bibnamefont {Kandala}}, \bibinfo
  {author} {\bibfnamefont {D.~C.}\ \bibnamefont {Mckay}}, \ and\ \bibinfo
  {author} {\bibfnamefont {J.~M.}\ \bibnamefont {Gambetta}},\ }\href {\doibase
  10.1103/PhysRevA.103.042605} {\bibfield  {journal} {\bibinfo  {journal}
  {Phys. Rev. A}\ }\textbf {\bibinfo {volume} {103}},\ \bibinfo {pages}
  {042605} (\bibinfo {year} {2021}{\natexlab{b}})}\BibitemShut {NoStop}%
\bibitem [{\citenamefont {Huang}\ \emph {et~al.}(2021)\citenamefont {Huang},
  \citenamefont {Broughton}, \citenamefont {Mohseni}, \citenamefont {Babbush},
  \citenamefont {Boixo}, \citenamefont {Neven},\ and\ \citenamefont
  {McClean}}]{huang2020power}%
  \BibitemOpen
  \bibfield  {author} {\bibinfo {author} {\bibfnamefont {H.-Y.}\ \bibnamefont
  {Huang}}, \bibinfo {author} {\bibfnamefont {M.}~\bibnamefont {Broughton}},
  \bibinfo {author} {\bibfnamefont {M.}~\bibnamefont {Mohseni}}, \bibinfo
  {author} {\bibfnamefont {R.}~\bibnamefont {Babbush}}, \bibinfo {author}
  {\bibfnamefont {S.}~\bibnamefont {Boixo}}, \bibinfo {author} {\bibfnamefont
  {H.}~\bibnamefont {Neven}}, \ and\ \bibinfo {author} {\bibfnamefont {J.~R.}\
  \bibnamefont {McClean}},\ }\href {\doibase 10.1038/s41467-021-22539-9}
  {\bibfield  {journal} {\bibinfo  {journal} {Nature Communications}\ }\textbf
  {\bibinfo {volume} {12}} (\bibinfo {year} {2021}),\
  10.1038/s41467-021-22539-9}\BibitemShut {NoStop}%
\bibitem [{\citenamefont {Shawe-Taylor}\ and\ \citenamefont
  {Cristianini}(2002)}]{shawe2002generalization}%
  \BibitemOpen
  \bibfield  {author} {\bibinfo {author} {\bibfnamefont {J.}~\bibnamefont
  {Shawe-Taylor}}\ and\ \bibinfo {author} {\bibfnamefont {N.}~\bibnamefont
  {Cristianini}},\ }\href {https://doi.org/10.1109/TIT.2002.802647} {\bibfield
  {journal} {\bibinfo  {journal} {IEEE Transactions on Information Theory}\
  }\textbf {\bibinfo {volume} {48}},\ \bibinfo {pages} {2721} (\bibinfo {year}
  {2002})}\BibitemShut {NoStop}%
\bibitem [{\citenamefont {Sch{\"o}lkopf}\ \emph {et~al.}(2001)\citenamefont
  {Sch{\"o}lkopf}, \citenamefont {Herbrich},\ and\ \citenamefont
  {Smola}}]{scholkopf2001generalized}%
  \BibitemOpen
  \bibfield  {author} {\bibinfo {author} {\bibfnamefont {B.}~\bibnamefont
  {Sch{\"o}lkopf}}, \bibinfo {author} {\bibfnamefont {R.}~\bibnamefont
  {Herbrich}}, \ and\ \bibinfo {author} {\bibfnamefont {A.~J.}\ \bibnamefont
  {Smola}},\ }in\ \href {https://doi.org/10.1007/3-540-44581-1_27} {\emph
  {\bibinfo {booktitle} {{Computational Learning Theory}}}}\ (\bibinfo
  {organization} {Springer},\ \bibinfo {year} {2001})\ pp.\ \bibinfo {pages}
  {416--426}\BibitemShut {NoStop}%
\bibitem [{\citenamefont {Nielsen}\ and\ \citenamefont
  {Chuang}(2010)}]{nielsen2010quantum}%
  \BibitemOpen
  \bibfield  {author} {\bibinfo {author} {\bibfnamefont {M.}~\bibnamefont
  {Nielsen}}\ and\ \bibinfo {author} {\bibfnamefont {I.}~\bibnamefont
  {Chuang}},\ }\href {https://books.google.com/books?id=j2ULnwEACAAJ} {\emph
  {\bibinfo {title} {Quantum Computation and Quantum Information: 10th
  Anniversary Edition}}}\ (\bibinfo  {publisher} {Cambridge University Press},\
  \bibinfo {year} {2010})\BibitemShut {NoStop}%
\bibitem [{\citenamefont {Cincio}\ \emph {et~al.}(2018)\citenamefont {Cincio},
  \citenamefont {Suba{\c{s}}{\i}}, \citenamefont {Sornborger},\ and\
  \citenamefont {Coles}}]{cincio2018learning}%
  \BibitemOpen
  \bibfield  {author} {\bibinfo {author} {\bibfnamefont {L.}~\bibnamefont
  {Cincio}}, \bibinfo {author} {\bibfnamefont {Y.}~\bibnamefont
  {Suba{\c{s}}{\i}}}, \bibinfo {author} {\bibfnamefont {A.~T.}\ \bibnamefont
  {Sornborger}}, \ and\ \bibinfo {author} {\bibfnamefont {P.~J.}\ \bibnamefont
  {Coles}},\ }\href@noop {} {\bibfield  {journal} {\bibinfo  {journal} {New
  Journal of Physics}\ }\textbf {\bibinfo {volume} {20}},\ \bibinfo {pages}
  {113022} (\bibinfo {year} {2018})}\BibitemShut {NoStop}%
\bibitem [{\citenamefont {Blum}\ and\ \citenamefont
  {Micali}(1984)}]{blum1984generate}%
  \BibitemOpen
  \bibfield  {author} {\bibinfo {author} {\bibfnamefont {M.}~\bibnamefont
  {Blum}}\ and\ \bibinfo {author} {\bibfnamefont {S.}~\bibnamefont {Micali}},\
  }\href {https://doi.org/10.1137/0213053} {\bibfield  {journal} {\bibinfo
  {journal} {SIAM journal on Computing}\ }\textbf {\bibinfo {volume} {13}},\
  \bibinfo {pages} {850} (\bibinfo {year} {1984})}\BibitemShut {NoStop}%
\bibitem [{\citenamefont {Markov}\ and\ \citenamefont
  {Saeedi}(2012)}]{markov2012constant}%
  \BibitemOpen
  \bibfield  {author} {\bibinfo {author} {\bibfnamefont {I.~L.}\ \bibnamefont
  {Markov}}\ and\ \bibinfo {author} {\bibfnamefont {M.}~\bibnamefont
  {Saeedi}},\ }\href {https://arxiv.org/abs/1202.6614} {\bibfield  {journal}
  {\bibinfo  {journal} {arXiv:1202.6614}\ } (\bibinfo {year}
  {2012})}\BibitemShut {NoStop}%
\bibitem [{\citenamefont {Watrous}(2000)}]{watrous2000succinct}%
  \BibitemOpen
  \bibfield  {author} {\bibinfo {author} {\bibfnamefont {J.}~\bibnamefont
  {Watrous}},\ }in\ \href {https://doi.org/10.1109/SFCS.2000.892141} {\emph
  {\bibinfo {booktitle} {Proceedings 41st Annual Symposium on Foundations of
  Computer Science}}}\ (\bibinfo {organization} {IEEE},\ \bibinfo {year}
  {2000})\ pp.\ \bibinfo {pages} {537--546}\BibitemShut {NoStop}%
\bibitem [{\citenamefont {Schlingemann}\ and\ \citenamefont
  {Werner}(2001)}]{schlingemann2001quantum}%
  \BibitemOpen
  \bibfield  {author} {\bibinfo {author} {\bibfnamefont {D.}~\bibnamefont
  {Schlingemann}}\ and\ \bibinfo {author} {\bibfnamefont {R.~F.}\ \bibnamefont
  {Werner}},\ }\href {https://doi.org/10.1103/PhysRevA.65.012308} {\bibfield
  {journal} {\bibinfo  {journal} {Physical Review A}\ }\textbf {\bibinfo
  {volume} {65}},\ \bibinfo {pages} {012308} (\bibinfo {year}
  {2001})}\BibitemShut {NoStop}%
\bibitem [{\citenamefont {Hein}\ \emph {et~al.}(2004)\citenamefont {Hein},
  \citenamefont {Eisert},\ and\ \citenamefont {Briegel}}]{hein2004multiparty}%
  \BibitemOpen
  \bibfield  {author} {\bibinfo {author} {\bibfnamefont {M.}~\bibnamefont
  {Hein}}, \bibinfo {author} {\bibfnamefont {J.}~\bibnamefont {Eisert}}, \ and\
  \bibinfo {author} {\bibfnamefont {H.~J.}\ \bibnamefont {Briegel}},\ }\href
  {https://doi.org/10.1103/PhysRevA.69.062311} {\bibfield  {journal} {\bibinfo
  {journal} {Physical Review A}\ }\textbf {\bibinfo {volume} {69}},\ \bibinfo
  {pages} {062311} (\bibinfo {year} {2004})}\BibitemShut {NoStop}%
\bibitem [{\citenamefont {Spall}(1992)}]{spall1992multivariate}%
  \BibitemOpen
  \bibfield  {author} {\bibinfo {author} {\bibfnamefont {J.}~\bibnamefont
  {Spall}},\ }\href {\doibase 10.1109/9.119632} {\bibfield  {journal} {\bibinfo
   {journal} {IEEE Transactions on Automatic Control}\ }\textbf {\bibinfo
  {volume} {37}},\ \bibinfo {pages} {332} (\bibinfo {year} {1992})}\BibitemShut
  {NoStop}%
\bibitem [{\citenamefont {Andersen}\ \emph {et~al.}()\citenamefont {Andersen},
  \citenamefont {Dahl},\ and\ \citenamefont {Vandenberghe}}]{cvxopt2021}%
  \BibitemOpen
  \bibfield  {author} {\bibinfo {author} {\bibfnamefont {M.}~\bibnamefont
  {Andersen}}, \bibinfo {author} {\bibfnamefont {J.}~\bibnamefont {Dahl}}, \
  and\ \bibinfo {author} {\bibfnamefont {L.}~\bibnamefont {Vandenberghe}},\
  }\href {https://cvxopt.org/} {\emph {\bibinfo {title} {{CVXOPT: Python
  Software for Convex Optimization. Version 1.2.6}}}}\BibitemShut {NoStop}%
\bibitem [{\citenamefont {Rafique}\ \emph {et~al.}(2018)\citenamefont
  {Rafique}, \citenamefont {Liu}, \citenamefont {Lin},\ and\ \citenamefont
  {Yang}}]{rafique2018non}%
  \BibitemOpen
  \bibfield  {author} {\bibinfo {author} {\bibfnamefont {H.}~\bibnamefont
  {Rafique}}, \bibinfo {author} {\bibfnamefont {M.}~\bibnamefont {Liu}},
  \bibinfo {author} {\bibfnamefont {Q.}~\bibnamefont {Lin}}, \ and\ \bibinfo
  {author} {\bibfnamefont {T.}~\bibnamefont {Yang}},\ }\href
  {https://arxiv.org/abs/1810.02060} {\bibfield  {journal} {\bibinfo  {journal}
  {arXiv:1810.02060}\ } (\bibinfo {year} {2018})}\BibitemShut {NoStop}%
\bibitem [{\citenamefont {Thekumparampil}\ \emph {et~al.}(2019)\citenamefont
  {Thekumparampil}, \citenamefont {Jain}, \citenamefont {Netrapalli},\ and\
  \citenamefont {Oh}}]{thekumparampil2019efficient}%
  \BibitemOpen
  \bibfield  {author} {\bibinfo {author} {\bibfnamefont {K.~K.}\ \bibnamefont
  {Thekumparampil}}, \bibinfo {author} {\bibfnamefont {P.}~\bibnamefont
  {Jain}}, \bibinfo {author} {\bibfnamefont {P.}~\bibnamefont {Netrapalli}}, \
  and\ \bibinfo {author} {\bibfnamefont {S.}~\bibnamefont {Oh}},\ }in\ \href
  {https://proceedings.neurips.cc/paper/2019/file/05d0abb9a864ae4981e933685b8b915c-Paper.pdf}
  {\emph {\bibinfo {booktitle} {Advances in Neural Information Processing
  Systems}}},\ Vol.~\bibinfo {volume} {32}\ (\bibinfo  {publisher} {Curran
  Associates, Inc.},\ \bibinfo {year} {2019})\BibitemShut {NoStop}%
\bibitem [{\citenamefont {Achlioptas}\ \emph {et~al.}(2002)\citenamefont
  {Achlioptas}, \citenamefont {McSherry},\ and\ \citenamefont
  {Sch{\"o}lkopf}}]{scholkopf2002sampling}%
  \BibitemOpen
  \bibfield  {author} {\bibinfo {author} {\bibfnamefont {D.}~\bibnamefont
  {Achlioptas}}, \bibinfo {author} {\bibfnamefont {F.}~\bibnamefont
  {McSherry}}, \ and\ \bibinfo {author} {\bibfnamefont {B.}~\bibnamefont
  {Sch{\"o}lkopf}},\ }\href
  {https://proceedings.neurips.cc/paper/2001/file/07cb5f86508f146774a2fac4373a8e50-Paper.pdf}
  {\bibfield  {journal} {\bibinfo  {journal} {Advances in neural information
  processing systems}\ }\textbf {\bibinfo {volume} {14}},\ \bibinfo {pages}
  {335} (\bibinfo {year} {2002})}\BibitemShut {NoStop}%
\bibitem [{\citenamefont {Kim}\ \emph {et~al.}(2006)\citenamefont {Kim},
  \citenamefont {Magnani},\ and\ \citenamefont {Boyd}}]{Kim2006kernels}%
  \BibitemOpen
  \bibfield  {author} {\bibinfo {author} {\bibfnamefont {S.-J.}\ \bibnamefont
  {Kim}}, \bibinfo {author} {\bibfnamefont {A.}~\bibnamefont {Magnani}}, \ and\
  \bibinfo {author} {\bibfnamefont {S.}~\bibnamefont {Boyd}},\ }in\ \href
  {\doibase 10.1145/1143844.1143903} {\emph {\bibinfo {booktitle} {Proceedings
  of the 23rd International Conference on Machine Learning}}},\ \bibinfo
  {series and number} {ICML '06}\ (\bibinfo  {publisher} {Association for
  Computing Machinery},\ \bibinfo {address} {New York, NY, USA},\ \bibinfo
  {year} {2006})\ pp.\ \bibinfo {pages} {465--472}\BibitemShut {NoStop}%
\bibitem [{\citenamefont {Gretton}\ \emph {et~al.}(2005)\citenamefont
  {Gretton}, \citenamefont {Bousquet}, \citenamefont {Smola},\ and\
  \citenamefont {Sch{\"o}lkopf}}]{Gretton2005kernels}%
  \BibitemOpen
  \bibfield  {author} {\bibinfo {author} {\bibfnamefont {A.}~\bibnamefont
  {Gretton}}, \bibinfo {author} {\bibfnamefont {O.}~\bibnamefont {Bousquet}},
  \bibinfo {author} {\bibfnamefont {A.}~\bibnamefont {Smola}}, \ and\ \bibinfo
  {author} {\bibfnamefont {B.}~\bibnamefont {Sch{\"o}lkopf}},\ }in\ \href
  {https://doi.org/10.1007/11564089_7} {\emph {\bibinfo {booktitle}
  {International conference on algorithmic learning theory}}}\ (\bibinfo
  {organization} {Springer},\ \bibinfo {year} {2005})\ pp.\ \bibinfo {pages}
  {63--77}\BibitemShut {NoStop}%
\bibitem [{\citenamefont {Hubregtsen}\ \emph {et~al.}(2021)\citenamefont
  {Hubregtsen}, \citenamefont {Wierichs}, \citenamefont {Gil-Fuster},
  \citenamefont {Derks}, \citenamefont {Faehrmann},\ and\ \citenamefont
  {Meyer}}]{hubregtsen2021training}%
  \BibitemOpen
  \bibfield  {author} {\bibinfo {author} {\bibfnamefont {T.}~\bibnamefont
  {Hubregtsen}}, \bibinfo {author} {\bibfnamefont {D.}~\bibnamefont
  {Wierichs}}, \bibinfo {author} {\bibfnamefont {E.}~\bibnamefont
  {Gil-Fuster}}, \bibinfo {author} {\bibfnamefont {P.-J.~H.}\ \bibnamefont
  {Derks}}, \bibinfo {author} {\bibfnamefont {P.~K.}\ \bibnamefont
  {Faehrmann}}, \ and\ \bibinfo {author} {\bibfnamefont {J.~J.}\ \bibnamefont
  {Meyer}},\ }\href@noop {} {\bibfield  {journal} {\bibinfo  {journal} {arXiv:2105.02276}\ } (\bibinfo {year} {2021})}\BibitemShut {NoStop}%
\bibitem [{\citenamefont {Srebro}\ and\ \citenamefont
  {Ben-David}(2006)}]{Srebro2006kernels}%
  \BibitemOpen
  \bibfield  {author} {\bibinfo {author} {\bibfnamefont {N.}~\bibnamefont
  {Srebro}}\ and\ \bibinfo {author} {\bibfnamefont {S.}~\bibnamefont
  {Ben-David}},\ }in\ \href {https://doi.org/10.1007/11776420_15} {\emph
  {\bibinfo {booktitle} {International Conference on Computational Learning
  Theory}}}\ (\bibinfo {organization} {Springer},\ \bibinfo {year} {2006})\
  pp.\ \bibinfo {pages} {169--183}\BibitemShut {NoStop}%
\bibitem [{\citenamefont {Jebara}(2004)}]{Jebara2006kernels}%
  \BibitemOpen
  \bibfield  {author} {\bibinfo {author} {\bibfnamefont {T.}~\bibnamefont
  {Jebara}},\ }\href {\doibase 10.1145/1015330.1015426} {\bibfield  {journal}
  {\bibinfo  {journal} {Proceedings of the 21st International Conference on
  Machine Learning}\ } (\bibinfo {year} {2004}),\
  10.1145/1015330.1015426}\BibitemShut {NoStop}%
\bibitem [{\citenamefont {Mcclean}\ \emph {et~al.}(2018)\citenamefont
  {Mcclean}, \citenamefont {Boixo}, \citenamefont {Smelyanskiy}, \citenamefont
  {Babbush},\ and\ \citenamefont {Neven}}]{McClean2018barren}%
  \BibitemOpen
  \bibfield  {author} {\bibinfo {author} {\bibfnamefont {J.}~\bibnamefont
  {Mcclean}}, \bibinfo {author} {\bibfnamefont {S.}~\bibnamefont {Boixo}},
  \bibinfo {author} {\bibfnamefont {V.}~\bibnamefont {Smelyanskiy}}, \bibinfo
  {author} {\bibfnamefont {R.}~\bibnamefont {Babbush}}, \ and\ \bibinfo
  {author} {\bibfnamefont {H.}~\bibnamefont {Neven}},\ }\href {\doibase
  10.1038/s41467-018-07090-4} {\bibfield  {journal} {\bibinfo  {journal}
  {Nature Communications}\ }\textbf {\bibinfo {volume} {9}} (\bibinfo {year}
  {2018}),\ 10.1038/s41467-018-07090-4}\BibitemShut {NoStop}%
\bibitem [{\citenamefont {Cerezo}\ \emph {et~al.}(2020)\citenamefont {Cerezo},
  \citenamefont {Sone}, \citenamefont {Volkoff}, \citenamefont {Cincio},\ and\
  \citenamefont {Coles}}]{Cerezo2020barren}%
  \BibitemOpen
  \bibfield  {author} {\bibinfo {author} {\bibfnamefont {M.}~\bibnamefont
  {Cerezo}}, \bibinfo {author} {\bibfnamefont {A.}~\bibnamefont {Sone}},
  \bibinfo {author} {\bibfnamefont {T.}~\bibnamefont {Volkoff}}, \bibinfo
  {author} {\bibfnamefont {L.}~\bibnamefont {Cincio}}, \ and\ \bibinfo {author}
  {\bibfnamefont {P.}~\bibnamefont {Coles}},\ }\href
  {https://arxiv.org/pdf/2001.00550.pdf} {\bibfield  {journal} {\bibinfo
  {journal} {arXiv:2001.00550v3}\ } (\bibinfo {year} {2020})}\BibitemShut
  {NoStop}%
\bibitem [{\citenamefont {Marrero}\ \emph {et~al.}(2021)\citenamefont
  {Marrero}, \citenamefont {Kieferova},\ and\ \citenamefont
  {Wiebe}}]{Marrero2021barren}%
  \BibitemOpen
  \bibfield  {author} {\bibinfo {author} {\bibfnamefont {C.}~\bibnamefont
  {Marrero}}, \bibinfo {author} {\bibfnamefont {M.}~\bibnamefont {Kieferova}},
  \ and\ \bibinfo {author} {\bibfnamefont {N.}~\bibnamefont {Wiebe}},\ }\href
  {\doibase 10.1103/PRXQuantum.2.040316} {\bibfield  {journal} {\bibinfo
  {journal} {PRX Quantum}\ }\textbf {\bibinfo {volume} {2}} (\bibinfo {year}
  {2021}),\ 10.1103/PRXQuantum.2.040316}\BibitemShut {NoStop}%
\bibitem [{\citenamefont {Motzoi}\ \emph {et~al.}(2009)\citenamefont {Motzoi},
  \citenamefont {Gambetta}, \citenamefont {Rebentrost},\ and\ \citenamefont
  {Wilhelm}}]{motzoi2009simple}%
  \BibitemOpen
  \bibfield  {author} {\bibinfo {author} {\bibfnamefont {F.}~\bibnamefont
  {Motzoi}}, \bibinfo {author} {\bibfnamefont {J.~M.}\ \bibnamefont
  {Gambetta}}, \bibinfo {author} {\bibfnamefont {P.}~\bibnamefont
  {Rebentrost}}, \ and\ \bibinfo {author} {\bibfnamefont {F.~K.}\ \bibnamefont
  {Wilhelm}},\ }\href {\doibase 10.1103/PhysRevLett.103.110501} {\bibfield
  {journal} {\bibinfo  {journal} {Phys. Rev. Lett.}\ }\textbf {\bibinfo
  {volume} {103}},\ \bibinfo {pages} {110501} (\bibinfo {year}
  {2009})}\BibitemShut {NoStop}%
\bibitem [{\citenamefont {Chow}\ \emph {et~al.}(2011)\citenamefont {Chow},
  \citenamefont {C\'orcoles}, \citenamefont {Gambetta}, \citenamefont
  {Rigetti}, \citenamefont {Johnson}, \citenamefont {Smolin}, \citenamefont
  {Rozen}, \citenamefont {Keefe}, \citenamefont {Rothwell}, \citenamefont
  {Ketchen},\ and\ \citenamefont {Steffen}}]{Chow2011}%
  \BibitemOpen
  \bibfield  {author} {\bibinfo {author} {\bibfnamefont {J.~M.}\ \bibnamefont
  {Chow}}, \bibinfo {author} {\bibfnamefont {A.~D.}\ \bibnamefont
  {C\'orcoles}}, \bibinfo {author} {\bibfnamefont {J.~M.}\ \bibnamefont
  {Gambetta}}, \bibinfo {author} {\bibfnamefont {C.}~\bibnamefont {Rigetti}},
  \bibinfo {author} {\bibfnamefont {B.~R.}\ \bibnamefont {Johnson}}, \bibinfo
  {author} {\bibfnamefont {J.~A.}\ \bibnamefont {Smolin}}, \bibinfo {author}
  {\bibfnamefont {J.~R.}\ \bibnamefont {Rozen}}, \bibinfo {author}
  {\bibfnamefont {G.~A.}\ \bibnamefont {Keefe}}, \bibinfo {author}
  {\bibfnamefont {M.~B.}\ \bibnamefont {Rothwell}}, \bibinfo {author}
  {\bibfnamefont {M.~B.}\ \bibnamefont {Ketchen}}, \ and\ \bibinfo {author}
  {\bibfnamefont {M.}~\bibnamefont {Steffen}},\ }\href {\doibase
  10.1103/PhysRevLett.107.080502} {\bibfield  {journal} {\bibinfo  {journal}
  {Phys. Rev. Lett.}\ }\textbf {\bibinfo {volume} {107}},\ \bibinfo {pages}
  {080502} (\bibinfo {year} {2011})}\BibitemShut {NoStop}%
\bibitem [{\citenamefont {Sundaresan}\ \emph {et~al.}(2020)\citenamefont
  {Sundaresan}, \citenamefont {Lauer}, \citenamefont {Pritchett}, \citenamefont
  {Magesan}, \citenamefont {Jurcevic},\ and\ \citenamefont
  {Gambetta}}]{Sundaresan2020}%
  \BibitemOpen
  \bibfield  {author} {\bibinfo {author} {\bibfnamefont {N.}~\bibnamefont
  {Sundaresan}}, \bibinfo {author} {\bibfnamefont {I.}~\bibnamefont {Lauer}},
  \bibinfo {author} {\bibfnamefont {E.}~\bibnamefont {Pritchett}}, \bibinfo
  {author} {\bibfnamefont {E.}~\bibnamefont {Magesan}}, \bibinfo {author}
  {\bibfnamefont {P.}~\bibnamefont {Jurcevic}}, \ and\ \bibinfo {author}
  {\bibfnamefont {J.~M.}\ \bibnamefont {Gambetta}},\ }\href {\doibase
  10.1103/PRXQuantum.1.020318} {\bibfield  {journal} {\bibinfo  {journal} {PRX
  Quantum}\ }\textbf {\bibinfo {volume} {1}},\ \bibinfo {pages} {020318}
  (\bibinfo {year} {2020})}\BibitemShut {NoStop}%
\bibitem [{\citenamefont {Kim}\ \emph {et~al.}(2021)\citenamefont {Kim},
  \citenamefont {Wood}, \citenamefont {Yoder}, \citenamefont {Merkel},
  \citenamefont {Gambetta}, \citenamefont {Temme},\ and\ \citenamefont
  {Kandala}}]{kim2021errormit}%
  \BibitemOpen
  \bibfield  {author} {\bibinfo {author} {\bibfnamefont {Y.}~\bibnamefont
  {Kim}}, \bibinfo {author} {\bibfnamefont {C.~J.}\ \bibnamefont {Wood}},
  \bibinfo {author} {\bibfnamefont {T.~J.}\ \bibnamefont {Yoder}}, \bibinfo
  {author} {\bibfnamefont {S.~T.}\ \bibnamefont {Merkel}}, \bibinfo {author}
  {\bibfnamefont {J.~M.}\ \bibnamefont {Gambetta}}, \bibinfo {author}
  {\bibfnamefont {K.}~\bibnamefont {Temme}}, \ and\ \bibinfo {author}
  {\bibfnamefont {A.}~\bibnamefont {Kandala}},\ }\href
  {https://arxiv.org/abs/2108.09197} {\bibfield  {journal} {\bibinfo  {journal}
  {arXiv:2108.09197}\ } (\bibinfo {year} {2021})}\BibitemShut {NoStop}%
\bibitem [{\citenamefont {Magesan}\ \emph {et~al.}(2011)\citenamefont
  {Magesan}, \citenamefont {Gambetta},\ and\ \citenamefont
  {Emerson}}]{Magesan2011}%
  \BibitemOpen
  \bibfield  {author} {\bibinfo {author} {\bibfnamefont {E.}~\bibnamefont
  {Magesan}}, \bibinfo {author} {\bibfnamefont {J.~M.}\ \bibnamefont
  {Gambetta}}, \ and\ \bibinfo {author} {\bibfnamefont {J.}~\bibnamefont
  {Emerson}},\ }\href {\doibase 10.1103/PhysRevLett.106.180504} {\bibfield
  {journal} {\bibinfo  {journal} {Phys. Rev. Lett.}\ }\textbf {\bibinfo
  {volume} {106}},\ \bibinfo {pages} {180504} (\bibinfo {year}
  {2011})}\BibitemShut {NoStop}%
\end{thebibliography}
%

\end{document}